\documentclass{aa}  

\usepackage{natbib}
\bibpunct{(}{)}{;}{a}{}{,} % to follow the A&A style
\usepackage{graphicx}
\usepackage{setspace}
\usepackage[colorlinks=True,allcolors=blue]{hyperref}
\usepackage{txfonts}
\usepackage{caption}
\usepackage{subcaption}
\usepackage{multirow}
\usepackage{lscape}
\usepackage{longtable}
\usepackage{rotating}
\begin{document}

   \title{Star-formation driven outflows in local dwarf galaxies as revealed from [CII] observations by \textit{Herschel}\thanks{\textit{Herschel} is an ESA space observatory with science instruments provided by European-led Principal Investigator consortia and with important participation from NASA.}}
   \authorrunning{M. Romano et al.}
   \titlerunning{Galactic outflows from [CII] observations in local dwarf galaxies}

\author{M. Romano\thanks{E-mail: michael.romano@ncbj.gov.pl}\inst{1,2}
\and
A. Nanni\inst{1,3}
\and
D. Donevski\inst{1,4,5}
\and
M. Ginolfi\inst{6}
\and
G.~C. Jones\inst{7}
\and
I. Shivaei\inst{8}
\and
Junais\inst{1}
\and
D. Salak\inst{9,10}
\and\\
P. Sawant\inst{1}
}

\institute{
National Centre for Nuclear Research, ul. Pasteura 7, 02-093 Warsaw, Poland
\and
INAF - Osservatorio Astronomico di Padova, Vicolo dell'Osservatorio 5, I-35122, Padova, Italy
\and
INAF - Osservatorio astronomico d'Abruzzo, Via Maggini SNC, 64100, Teramo, Italy
\and
SISSA, Via Bonomea 265, Trieste, Italy
\and
IFPU - Institute for fundamental physics of the Universe, Via Beirut 2, 34014 Trieste, Italy
\and
Dipartimento di Fisica e Astronomia, Università di Firenze, Via G. Sansone 1, 50019, Sesto Fiorentino (Firenze), Italy
\and
Department of Physics, University of Oxford, Denys Wilkinson Building, Keble Road, Oxford OX1 3RH, UK
\and
Steward Observatory, University of Arizona, 933 North Cherry Avenue, Tucson, AZ 85721, USA
\and
Institute for the Advancement of Higher Education, Hokkaido University, Kita 17 Nishi 8, Kita-ku, Sapporo, Hokkaido 060-0817,
Japan
\and
Department of Cosmosciences, Graduate School of Science,
Hokkaido University, Kita 10 Nishi 8, Kita-ku, Sapporo, Hokkaido
060-0817, Japan
}

\date{Accepted 16 June 2023}

% \abstract{}{}{}{}{} 
% 5 {} token are mandatory
 
\abstract
{We characterize the physical properties of star-formation driven outflows in a sample of 29 local dwarf galaxies drawn from the Dwarf Galaxy Survey. We make use of \textit{Herschel}/PACS archival data to search for atomic outflow signatures in the wings of individual [CII] 158~$\mu$m spectra and in their stacked line profile. We find a clear excess of emission in the high-velocity tails of 11 sources which can be explained with an additional broad component (tracing the outflowing gas) in the modeling of their spectra. The remaining objects are likely hosts of weaker outflows that can still be detected in the average stacked spectrum. In both cases, we estimate the atomic mass outflow rates which result to be comparable with the star-formation rates of the galaxies, implying mass-loading factors (i.e., outflow efficiencies) of the order of unity. Outflow velocities in all the 11 galaxies with individual detections are larger than (or compatible with) the escape velocities of their dark matter halos, with an average fraction of 40\% of gas escaping into the intergalactic medium (IGM). Depletion timescales due to outflows are lower than those due to gas consumption by star formation in most of our sources, ranging from hundred million to a few billion years. From the energetic point of view, our outflows are mostly consistent with momentum-driven winds generated by the radiation pressure of young stellar populations on dust grains, although the energy-driven scenario is not excluded if considering a coupling efficiency up to 20\% between the energy injected by supernova (SN) and the interstellar medium. Overall, our results suggest that, despite their low efficiencies, galactic outflows can regulate the star formation history of dwarf galaxies. Specifically, they are able to enrich with metals the circumgalactic medium of these sources, bringing on average a non-negligible amount of gas into the IGM, where it will not be available anymore for new star formation. Our findings are suitable for tuning chemical evolution models attempting to describe the physical processes shaping the evolution of dwarf galaxies.} 
  
\keywords{Galaxies: dwarf - Galaxies: evolution - Galaxies: ISM - Galaxies: starburst - ISM: jets and outflows}

\maketitle
%
%-------------------------------------------------------------------

\section{Introduction}
The series of processes involved in the formation and evolution of galaxies across cosmic time is still far from being fully understood. What we undoubtedly know is that the amount of cold gas inside a galaxy is the main master of its fate, being the fuel of star formation. Indeed, gas can cool down and condense to form stars which, during their lifetime, alter the state and growth of a galaxy (see e.g., \citealt{Tumlinson17,Peroux20,Tacconi20} for reviews). In particular, massive stars are able to inject energy and momentum in their surroundings through stellar winds and supernova (SN) explosions, modifying the properties of the interstellar medium (ISM) by enriching it with heavy elements and sweeping the gas out of the galaxy via fast and powerful outflows (e.g., \citealt{Murray05,Veilleux05,Hopkins12,Erb15,Heckman15}). At the same time, SN can trigger interstellar turbulence regulating the star-formation rate (SFR) of a galaxy which, in turn, plays a key role in providing new sources of radiation and feedback (e.g., \citealt{Faucher13,Martizzi16,Orr18,Ostriker22}). Along with other mechanisms, such as radiation pressure (e.g., \citealt{Thompson05,Veilleux05,Murray11,Hopkins12}), cosmic rays (e.g., \citealt{Samui10,Hanasz13,Salem14}), and active galactic nuclei (AGNs; e.g., \citealt{Murray05,Faucher12,Harrison14,Rupke17}), stellar feedback rules the baryon cycle in galaxies, and it is essential in galaxy evolution models and simulations based on the Lambda-Cold Dark Matter ($\Lambda-$CDM) framework in order to reproduce their observational properties (e.g., \citealt{Springel05,Vogelsberger14,Schaye15,Pillepich18}). For instance, feedback from SN and AGNs is invoked to suppress star formation efficiency in low- and high-mass galaxies, respectively, lessening the discrepancies between the observed shape of the galaxy luminosity function and the predicted dark matter halo mass function (e.g., \citealt{Silk12,Behroozi13}). It is also needed to explain the co-evolution of central supermassive black holes with their host galaxies (e.g., \citealt{Tremaine02,Kormendy13,King15}), and many fundamental scaling relations, such as the mass-metallicity (e.g., \citealt{Mannucci09,Lilly13,Kashino16,Lian18,Curti20}) or Tully-Fisher relations (e.g., \citealt{McGaugh12,Somerville15}).  

Dwarf starburst galaxies (with stellar mass $M_{*}<10^{10}~M_{\odot}$; e.g., \citealt{Sartori15,McCormick18,Marasco22}) represent the ideal targets to investigate the impact of stellar feedback on galaxy evolution. In these sources, galactic winds are thought to be driven by the radiation from young stellar populations and SN explosions and, because of the shallow gravitational potential wells of their hosts, they can be much more effective than in high-mass galaxies in carrying large amount of metals and dust into the circumgalactic (or even intergalactic) medium (CGM or IGM; e.g., \citealt{Gnedin10,Booth12,Cote15,Schaye15,Dave17,Christensen18}). Moreover, dwarf low-metallicity galaxies are thought to be analogs of high-$z$ sources (e.g., \citealt{Patej15,Izotov21,Shivaei22}), allowing us to provide valuable insights on the processes shaping the evolution of their counterparts in the early universe. A key parameter for characterizing the power and efficiency of galactic outflows is the ratio between the rate at which the material in the ISM is expelled out of the galaxy (i.e., the mass outflow rate $\dot{M}_\mathrm{out}$) and its SFR, also known as the mass-loading factor ($\eta=\dot{M}_\mathrm{out}$/SFR).

On the theoretical side, predictions on the outflow efficiency can be obtained through cosmological hydrodynamical simulations. These models are able to simulate a large number of galaxies at different cosmic times, trying to reproduce the overall properties of the observed universe (e.g., \citealt{Dave11,Vogelsberger14,Muratov15,Nelson19}). However, they do not have enough resolution to unveil the physical processes of the stellar feedback taking place at the smallest scales, for which zoom-in simulations and semi-analytical models are needed (e.g., \citealt{Somerville08,Hopkins14,Cote15,Kim18}). In addition, chemical evolution models are another useful tool to investigate the need of galactic outflows in shaping the baryon cycle of galaxies (e.g., \citealt{Cote16,Nanni20,Galliano21}). As a matter of fact, all these models typically require large values of the mass-loading factor ($\eta>1$) in order to reproduce the observational properties of low-mass galaxies, although with a large scatter in their predictions. For instance, \cite{Nanni20} made use of the One-Zone Model for the Evolution of GAlaxies (\texttt{OMEGA}; \citealt{Cote17}) to model the chemical evolution of local low-metallicity dwarf galaxies from the Dwarf Galaxy Survey (DGS; \citealt{Madden13,Madden14}), along with high-redshift Lyman Break Galaxies (LBGs). They found that galactic outflows are crucial in order to reproduce e.g., the observed relatively low content of dust to stars in older sources, and they are more efficient than grain destruction by Type II SN in removing dust from the ISM of both local and high-redshift galaxies. However, the mass-loading factors they provide as input for their models depend on the initial gas mass in the galaxies, and they could span a very broad range of values (i.e., $\eta\sim0-80$) to cover the whole parameter space of their sources. Therefore, observational constraints on this parameter are pivotal for a better description of stellar feedback, as well as to disentangle different mechanisms of gas and dust production/destruction into the ISM of galaxies. 

Unsurprisingly, the mass-loading factor is challenging to be constrained as it depends on assumptions on the outflow physical size, geometry and composition (e.g., its temperature, density or chemistry), and it could be thus subject to many uncertainties (e.g., \citealt{Veilleux05,Maiolino12,Chisholm17,Fluetsch19,Lutz20}). Furthermore, outflows are composed of multiple gas phases (i.e., hot, warm, and cold) which require different techniques and instruments for an in-depth investigation. Such observations span the whole electromagnetic spectrum, ranging from the rest-frame UV/optical emission for the warm ionized, and cold atomic and molecular gas (e.g., \citealt{Contursi13,Heckman15,Gonzalez17,Fluetsch19,Herrera19,Concas22}), to the X-ray emission for the hot phase (e.g., \citealt{Heckman95,Ott05,Tombesi15,McQuinn18}). The expelled material is typically detected as a blueshift in the profile of UV/optical and far-infrared (FIR) absorption lines with respect to the systemic velocity of the galaxy (e.g., \citealt{Pettini02,Erb12,Veilleux13,Heckman15,Falgarone17,Gonzalez17,Talia17,Riechers21,Calabro22}), or by searching for a broad emission (that is supposed to trace the outflow) on top of a narrow component (related to the virial motion of stars) in the spectrum of observed emission lines (e.g., \citealt{Rupke13,Arribas14,Forster14,Janssen16,Herrera19,Marasco22}). 

In particular, the fine-structure transition of C$^{+}$ at 158~$\mu$m (hereafter, [CII]) has been proved to be suitable for characterizing a number of properties of the ISM of both local and high-redshift sources. This represents the brightest emission line in the rest-frame FIR spectra of star-forming galaxies (SFGs), being one of the major coolant of their ISM (e.g., \citealt{Stacey91,Carilli13}). The bulk of its emission comes from neutral atomic gas arising in photo-dissociation regions (PDRs) surrounding young stars \citep{Hollenbach99} but, given its low ionization potential (11.3 eV, compared to the 13.6 eV of neutral hydrogen), it can also be emitted by the partly ionized (e.g., \citealt{Cormier12,Pineda14}) and molecular medium (e.g., \citealt{Zanella18,Madden20}). Furthermore, [CII] has been recently used to infer the neutral hydrogen (HI) gas mass in galaxies out to $z\sim6$ both with observations and simulations (e.g., \citealt{Heintz21,Vizgan22}). High-velocity outflows have been detected
in the broad wings of the [CII] line profile in individual high-redshift luminous quasars (QSOs; e.g., \citealt{Maiolino12,Cicone15}) and in normal SFGs at $z > 4$ (e.g., \citealt{Ginolfi20,Herrera21}) thanks to the IRAM Plateau de Bure Interferometer and ALMA observations, respectively. In the local universe, \textit{Herschel} data have been exploited to trace atomic (and possibly molecular) outflows in the broad [CII] wings of ultra-luminous infrared galaxies (ULIRGs; e.g., \citealt{Janssen16}), as well.

The aim of this work is to further investigate the importance of stellar feedback in the evolution of galaxies by constraining the efficiency of galactic outflows in local dwarf sources. To do that, we make use of archival spectroscopic [CII] observations as collected by \textit{Herschel}/PACS in a sample of local dwarf galaxies drawn from the DGS. We explore both the individual and average properties of outflows in these sources, trying to characterize their efficiency, origin, and impact on their host galaxies and external environment. Our results will be used as input in state-of-the-art chemical evolution models in order to better understand the processes involved in the cycle of baryons into galaxies at different cosmic times (Nanni et al. in prep.). 

The paper is structured as follows. In Section \ref{sec:sample}, we provide a description of the dwarf galaxy sample used in our analysis. The data retrieving and reduction are described in Section \ref{sec:data}. In Section \ref{sec:analysis}, we describe the methods used to identify galactic outflows in individual galaxies, and via stacking of their spectra and data cubes. We present our results in Section \ref{sec:discussion}, including estimates of the outflow efficiencies and depletion timescales, as well as discussions on their ability to enrich the environment of their hosts, and their powering mechanisms. Summary and conclusions are reported in Section \ref{sec:summary}. Throughout this work, we adopt a $\Lambda-$CDM cosmology with $H_0 = 70~\mathrm{km~s^{-1}~Mpc^{-1}}$, $\mathrm{\Omega_m = 0.3}$ and $\mathrm{\Omega_\Lambda = 0.7}$. We also assume a \cite{Chabrier03} initial mass function.

%Why is it important to constrain outflows in these galaxies (refer to theoretical works)?\\
%Check \cite{Concas22} for ionized outflow in local SFGs and its low contribution to the total outflow (AND FOR DISCUSSION ON BROAD WINGS AS TRACERS OF OUTFLOWING GAS).\\
%Use [OIII]88~$\mu$m to trace ionized gas \citep{Carilli13,DeLooze14,Cormier15}.\\
%\cite{Contursi13} for comparison with H$\alpha$ as a tracer of ionized outflow.\\
%Check \cite{Manzano19} for escape velocity calculations.\\
%\cite{Marasco22} for new results on ionized outflow in dwarf galaxies including DGS.

\section{Sample description}\label{sec:sample}
The Dwarf Galaxy Survey (DGS\footnote{An overview of the survey, the sample selection, and the information available for each galaxy can be found at \url{https://irfu.cea.fr/Pisp/diane.cormier/dgsweb/index.html}.}; \citealt{Madden13,Madden14}) collected \textit{Herschel} observations of 48 low-metallicity ($7.14<\mathrm{12+log(O/H)<8.43}$, ranging from 1/50 to 1/2 $\mathrm{Z_{\odot}}$) local dwarf galaxies, with distances smaller than 200~Mpc and stellar masses in the range $\mathrm{log(M_{*}/M_{\odot})\sim6-10}$. The DGS sample was initially selected from several surveys targeting emission-line and blue compact dwarf galaxies, including the Hamburg/SAO survey and the First and Second Byurakan Surveys (e.g., \citealt{Markarian83,Izotov91,Ugryumov03}), with the aim of maximizing the number of sources with available multi-wavelength ancillary data over a wide metallicity range. The PACS \citep{Poglitsch10} and SPIRE \citep{Griffin10} instruments on board the \textit{Herschel} Space Observatory \citep{Pilbratt10} provided a full photometric and spectroscopic coverage of the FIR emission, retrieving information on the different phases of the ISM, as well as on the dust properties and star formation mechanisms in these galaxies (e.g., \citealt{Remy-Ruyer13,Cormier15,Cormier19}).

The sample is divided in 37 compact objects (fitting within the PACS footprint; see Sect. \ref{sec:data}) and 11 more extended sources which are also the closest to our galaxy. As the latter were observed only partially with the PACS spectrometer, we decided to exclude them from our analysis, focusing on the compact sample whose galaxies were entirely covered by the PACS pointings. Additionally, two faint sources were dropped from the PACS spectroscopy program because of time constraints \citep{Cormier15}, resulting in 35 compact galaxies.

\subsection{Physical parameters estimation}
\begin{figure}
    \begin{center}
	\includegraphics[width=\columnwidth]{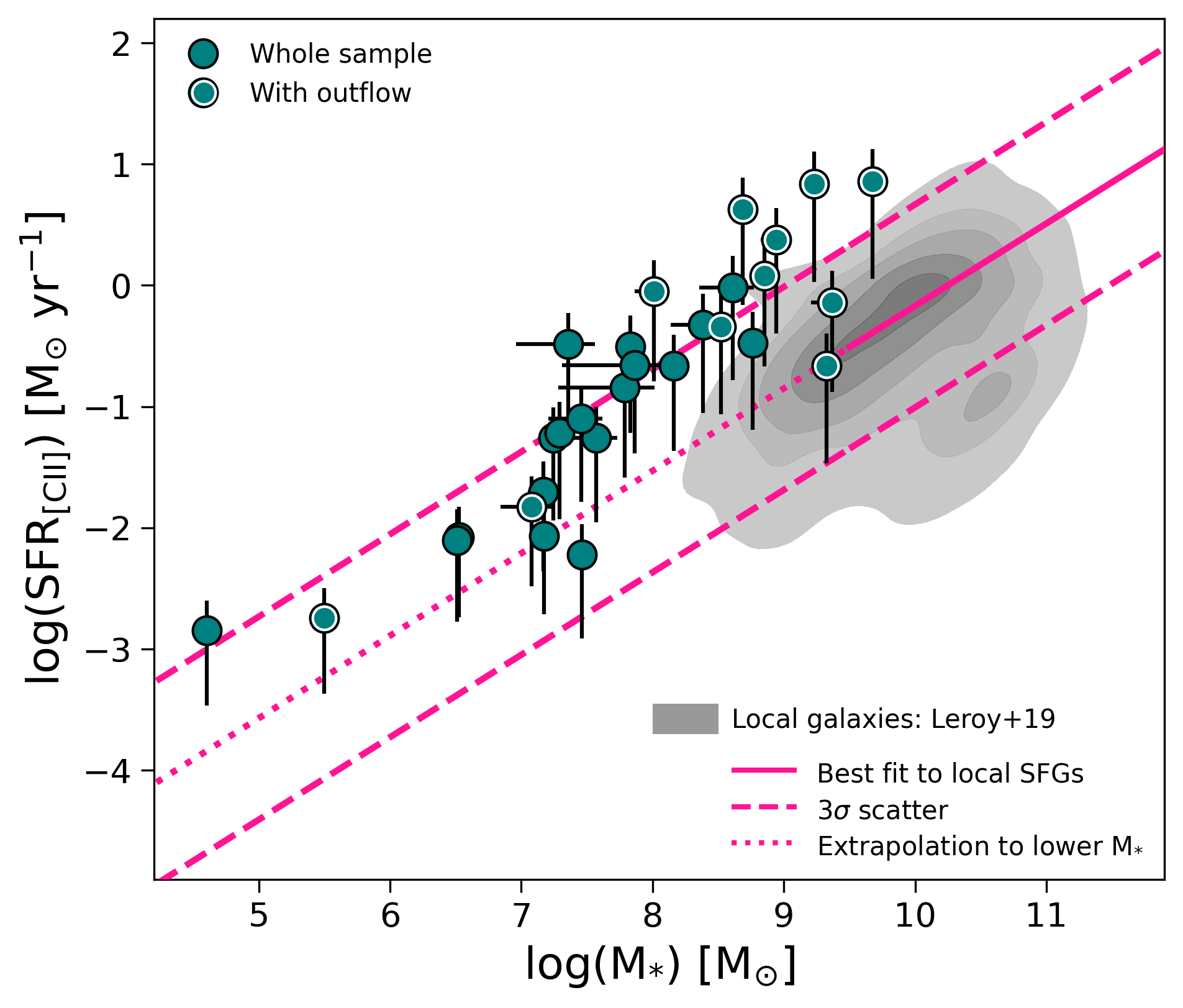}
	\end{center}
    \caption{SFR-$\mathrm{M_{*}}$ diagram for all the sources in our sample (teal circles). Grey contours show the distribution of nearby galaxies from \cite{Leroy19}. Here, contours increase in step of 20\%, with the lowest one including 90\% of the local sample. The pink solid line represents the best-fit relation by \cite{Leroy19} to local SFGs (selected in the range $\mathrm{log(M_{*}/M_{\odot})=9.5-11}$), probing the local star-forming main-sequence. We also show the extrapolation of such relation to the lower stellar mass regime covered by our galaxy sample (pink dotted line). Pink dashed lines are the $3\sigma$ scatter associated to the relation. Circles highlighted in white represent galaxies from our sample with individual outflow detections.}
    \label{fig:ms}
\end{figure}
The stellar masses and SFRs for the DGS galaxies are presented in \cite{Madden13,Madden14}. However, such quantities are not available for all the observed sources and they were retrieved with different methods, e.g., SFRs were obtained both from the IR luminosity or (when IR data were not present) from H$\alpha$ and H$\beta$ lines. In order to have more homogeneously derived physical quantities, we used the latest version of the Code Investigating GALaxy Emission (CIGALE; \citealt{Burgarella05,Noll09,Boquien19}) to recompute the physical parameters of our galaxies through spectral energy distribution (SED)-fitting. 

\begin{table*}
\caption{Parameters used in CIGALE for modeling the SEDs of our galaxies.}
\begin{center}
\begin{tabular}{l c c c c}
\hline
\hline
Parameters & Values & Unit & Description\\
\hline
\multicolumn{4}{c}{\textit{Delayed star-formation history}}\\
$\tau_\mathrm{main}$ & 25, 50, 100, 250 & Myr & e-folding time of main stellar population\\
Age$_\mathrm{main}$ & 101 log values in [50, 1260] & Myr & Age of the main stellar population\\
$f_\mathrm{burst}$ & No burst & - & Burst\\
\hline
\multicolumn{4}{c}{\textit{Stellar emission}}\\
SSP & \cite{Bruzual03} & - & Single stellar population\\
IMF & \cite{Chabrier03} & - & Initial mass function\\
Z & 0.0004, 0.004, 0.008 & - & Metallicity\\
Age separation & 10 & Myr & Age difference between old and young stars\\
\hline
\multicolumn{4}{c}{\textit{Nebular emission}}\\
log$U$ & \cite{Cormier19} & - & Ionization parameter\\
Z$_\mathrm{gas}$ & \cite{Madden13} & - & Gas metallicity\\
$n_\mathrm{e}$ & 10, 100, 1000 & cm$^{-3}$ & Electron density\\
- & 100 & km~s$^{-1}$ & Lines width\\
\hline
\multicolumn{4}{c}{\textit{Dust attenuation}}\\
- & \cite{Calzetti00} & - & Dust attenuation law\\
E\_BV\_lines & 101 log values in [0.001, 1] & - & Colour excess for old and young stars\\
uv\_bump\_ampl & 0.0 & - & Amplitude of the UV bump\\
powerlaw\_slope & 0.0 & - & Powerlaw slope modifying attenuation curve\\
\hline
\multicolumn{4}{c}{\textit{Dust emission}}\\
- & \cite{Draine14} & - & Dust emission models\\
$q_\mathrm{PAH}$ & \cite{Remy-Ruyer15} & - & Mass fraction of PAH\\
$U_\mathrm{min}$ & \cite{Remy-Ruyer15} & Habing & Minimum radiation field\\
$\alpha$ & \cite{Remy-Ruyer15} & - & Powerlaw slope d$U$/d$M_d\propto U{^\alpha}$\\
$\gamma$ & 0.1, 0.3, 0.5, 0.7, 0.9 & - & Illuminated fraction\\
\hline
\end{tabular}
\tablefoot{We did not include any AGN emission.}
\end{center}
\label{tab:CIGALE}
\end{table*}

We adopted the same photometry as in \cite{Burgarella20}, which includes UV and optical fluxes from the NASA/IPAC Extragalactic Database\footnote{The NASA/IPAC Extragalactic Database (NED) is operated by the Jet Propulsion Laboratory, California Institute of Technology,
under contract with the National Aeronautics and Space Administration.} (NED), 2MASS $J$, $H$, and $K$ near-IR bands, \textit{Spitzer} IRAC, IRAS, and WISE fluxes for the mid-IR, as well as \textit{Herschel} PACS and SPIRE coverage of the FIR regime. CIGALE modules and input parameters are also taken from \cite{Burgarella20} for local low-metallicity galaxies, with the exception of the following additional quantities that have been fixed in the fit, based on previous works on DGS galaxies. As the gas-phase metallicity ($\mathrm{Z_{gas}}$) in the DGS sample spans more than an order of magnitude, we set it for each source to the closest value from \cite{Madden13}. Based on that, we made multiple runs fixing each time the stellar metallicity (Z) to the available input values lower than $\mathrm{Z_{gas}}$ (e.g., \citealt{Lian18,Fraser22}), finally taking the one providing the best fit to the data. Furthermore, \cite{Remy-Ruyer15} adopted semi-empirical models by \cite{Galliano11} to fit the observed dust SED of DGS galaxies, recovering information on the formation and evolution of dust in these low-metallicity sources. In particular, we took advantage of their predictions on the mass fraction of polycyclic aromatic hydrocarbon ($q_\mathrm{PAH}$), the minimum value of the radiation field ($U_\mathrm{min}$), and the power-law slope of the distribution of its intensity per dust mass ($\alpha$, being $dU/dM_d\propto U^{\alpha})$. Constraints on the ionization parameter ($U$) were provided by \cite{Cormier19} who made use of the spectral synthesis code Cloudy \citep{Ferland17} to model the ISM phases of the DGS sources, trying to reproduce at the same time the corresponding infrared luminosities estimated by \cite{Remy-Ruyer15}. For each of these quantities, we took the values closer to the corresponding model grids, and we gave those as input in CIGALE. We added more freedom to each fit by letting these input parameters to vary within their uncertainties, reaching an average reduced $\chi^2\sim1.9$. %put in CIGALE the closest number to that estimated in these works, letting it to vary between its preceding and following value to add more freedom to the fit. 
Final modules and input parameters are reported in Table \ref{tab:CIGALE}.

We thus obtained, in a consistent way, new estimates of $\mathrm{M_{*}}$ for all the galaxies in our sample. Our results are in agreement with previous findings by \cite{Burgarella20}, although systematically lower (up to $\sim1~$dex in a few cases, as similarly found in \citealt{Nanni20}) than original results by \cite{Madden13}. In their work, \cite{Madden13} computed DGS stellar masses based on \textit{Spitzer} IRAC 3.6 and 4.5~$\mu$m bands, assuming a constant near-IR mass-to-light ratio, without accounting for any dependence on metallicity and age \citep{Eskew12,Wen13}. Here, we made a detailed SED-fitting of each galaxy by modeling their emission from UV/optical to FIR wavelengths, individually tuning their gas and dust properties from previous analysis and observations (see Table \ref{tab:CIGALE}). Clearly, the adopted SFH and/or IMF could have an impact on the estimated $\mathrm{M_{*}}$ (see e.g., discussion in Appendix B.2 of \citealt{Galliano21}). Before picking out the delayed one, \cite{Burgarella20} explored different SFHs in CIGALE to reproduce the SEDs of the DGS galaxies, finding no significant improvement of the fits. Furthermore, recent results by \cite{Motino21} for local dwarf galaxies (including a few DGS sources) suggest no evidence for old stellar populations for the majority of their sample, in contrast with the SFHs assumed by \cite{Galliano21} that are modeled on two (old and young) stellar populations, and more in agreement with those adopted in this work (see also Sect. \ref{subsec:outflow_efficiency}). Finally, we note that the mass-to-light ratio used by \cite{Madden13} to compute $\mathrm{M_{*}}$ is calibrated on a \cite{Salpeter55} IMF. By applying a systematic scaling of a factor $\sim0.6$ to our stellar masses (to convert from \citealt{Chabrier03} to \citealt{Salpeter55} IMF; e.g., \citealt{Madau14,Galliano21}) would alleviate the tension between our estimates and those by \cite{Madden13}. For all these reasons, we are confident about our new $\mathrm{M_{*}}$ results, although further investigation is needed to reduce the difference in the stellar mass computation among various methods. 

About SFRs, we decided to compute them from observations. In particular, we adopted the prescription presented by \cite{DeLooze14}, that was specifically calibrated on the DGS galaxies (i.e., $\mathrm{log(SFR)} = (-5.73\pm0.32) + (0.80\pm0.05) \times \mathrm{log}(L_\mathrm{[CII]})$), using the observed [CII] luminosity as a tracer of the total SFR (see Sect. \ref{subsec:individual_out}). On average, we found that our new estimates of the SFRs from [CII] (i.e., SFR$_\mathrm{[CII]}$) are in agreement with those obtained with CIGALE within the uncertainties. We also note here that SFRs used to calibrate the [CII]-SFR relation by \cite{DeLooze14} are not the same estimated by \cite{Madden13}. In particular, \cite{DeLooze14} used the GALEX FUV and MIPS 24~$\mu$m fluxes to probe the dust unobscured and obscured star formation, respectively, both available for 32 out of 48 DGS sources.

\begin{table*}
\caption{Physical properties of our sample of galaxies.}
\label{tab:summary}
\begin{center}
\begin{spacing}{1.25}
\begin{tabular}{l c c c c c}
\hline
\hline
Source & $\mathrm{z_{[CII]}}$ & $\mathrm{log(M_{*})}$ & log(SFR$_\mathrm{[CII]}$) & $\mathrm{log(M_{H2,TOT})}$ & $\mathrm{12+log(O/H)}$\\
 & & $\mathrm{[M_{\odot}]}$ & $\mathrm{[M_{\odot}~yr^{-1}]}$ & $\mathrm{[M_{\odot}]}$ & \\
 (1) & (2) & (3) & (4) & (5) & (6)\\
\hline
Haro2 & 0.0049 & $8.01^{+0.11}_{-0.15}$ & $-0.05^{+0.26}_{-0.74}$ & $9.01^{+0.15}_{-0.23}$ & $8.23\pm0.03$\\
Haro3 & 0.0032 & $8.52^{+0.07}_{-0.09}$ & $-0.34^{+0.26}_{-0.72}$ & $8.66^{+0.15}_{-0.23}$ & $8.28\pm0.01$\\
Haro11 & 0.0207 & $9.23^{+0.06}_{-0.07}$ & $0.84^{+0.27}_{-0.81}$ & $10.08^{+0.15}_{-0.23}$ & $8.36\pm0.01$\\
He2-10 & 0.0029 & $8.86^{+0.07}_{-0.09}$ & $0.08^{+0.26}_{-0.75}$ & $9.16^{+0.15}_{-0.23}$ & $8.43\pm0.01$\\
HS0052+2536 & 0.0462 & $8.61^{+0.16}_{-0.26}$ & $-0.02^{+0.26}_{-0.76}$ & $9.05^{+0.16}_{-0.25}$ & $8.07\pm0.20$\\
HS1222+3741 & 0.0405 & $9.33^{+0.03}_{-0.03}$ & $-0.66^{+0.27}_{-0.81}$ & $8.26^{+0.18}_{-0.32}$ & $7.79\pm0.01$\\ 
HS1304+3529 & 0.0162 & $7.36^{+0.20}_{-0.40}$ & $-0.49^{+0.26}_{-0.73}$ & $8.48^{+0.15}_{-0.24}$ & $7.93\pm0.10$\\ 
HS1330+3651 & 0.0165 & $8.38^{+0.16}_{-0.24}$ & $-0.33^{+0.26}_{-0.73}$ & $8.67^{+0.15}_{-0.24}$ & $7.98\pm0.10$\\ 
HS1442+4250 & 0.0022 & $7.46^{+0.07}_{-0.09}$ & $-2.22^{+0.25}_{-0.69}$ & $6.37^{+0.17}_{-0.29}$ & $7.60\pm0.01$\\ 
HS2352+2733 & 0.0273 & $7.79^{+0.23}_{-0.51}$ & $-0.84^{+0.26}_{-0.74}$ & $8.05^{+0.17}_{-0.28}$ & $8.40\pm0.20$\\ 
IZw18 & 0.0025 & $6.52^{+0.05}_{-0.05}$ & $-2.08^{+0.25}_{-0.66}$ & $6.55^{+0.16}_{-0.25}$ & $7.14\pm0.01$\\ 
IIZw40 & 0.0026 & $7.83^{+0.07}_{-0.08}$ & $-0.50^{+0.26}_{-0.71}$ & $8.46^{+0.15}_{-0.23}$ & $8.23\pm0.01$\\ 
Mrk153 & 0.0080 & $8.16^{+0.05}_{-0.06}$ & $-0.66^{+0.26}_{-0.70}$ & $8.27^{+0.15}_{-0.23}$ & $7.86\pm0.04$\\ 
Mrk209 & 0.0010 & $7.17^{+0.02}_{-0.03}$ & $-2.07^{+0.25}_{-0.64}$ & $6.56^{+0.15}_{-0.23}$ & $7.74\pm0.01$\\ 
Mrk930 & 0.0183 & $8.94^{+0.09}_{-0.12}$ & $0.38^{+0.26}_{-0.77}$ & $9.52^{+0.15}_{-0.24}$ & $8.03\pm0.01$\\ 
Mrk1089 & 0.0134 & $8.68^{+0.09}_{-0.11}$ & $0.62^{+0.26}_{-0.79}$ & $9.82^{+0.15}_{-0.23}$ & $8.10\pm0.08$\\ 
Mrk1450 & 0.0032 & $7.25^{+0.07}_{-0.08}$ & $-1.26^{+0.25}_{-0.68}$ & $7.54^{+0.15}_{-0.24}$ & $7.84\pm0.01$\\ 
SBS0335-052 & 0.0136 & $7.29^{+0.06}_{-0.08}$ & $-1.22^{+0.26}_{-0.71}$ & $7.59^{+0.16}_{-0.27}$ & $7.25\pm0.01$\\ 
SBS1159+545 & 0.0121 & $7.57^{+0.16}_{-0.26}$ & $-1.26^{+0.26}_{-0.70}$ & $7.54^{+0.16}_{-0.25}$ & $7.44\pm0.01$\\ 
SBS1211+540 & 0.0030 & $6.51^{+0.07}_{-0.09}$ & $-2.10^{+0.25}_{-0.67}$ & $6.52^{+0.16}_{-0.26}$ & $7.58\pm0.01$\\ 
SBS1249+493 & 0.0248 & $7.86^{+0.24}_{-0.55}$ & $-0.66^{+0.26}_{-0.73}$ & $8.27^{+0.16}_{-0.25}$ & $7.68\pm0.02$\\ 
SBS1415+437 & 0.0021 & $7.17^{+0.09}_{-0.12}$ & $-1.70^{+0.25}_{-0.66}$ & $7.00^{+0.15}_{-0.23}$ & $7.55\pm0.01$\\ 
SBS1533+574 & 0.0114 & $8.76^{+0.04}_{-0.04}$ & $-0.48^{+0.26}_{-0.71}$ & $8.49^{+0.15}_{-0.23}$ & $8.05\pm0.01$\\ 
UGC4483 & 0.0006 & $5.50^{+0.06}_{-0.06}$ & $-2.74^{+0.25}_{-0.62}$ & $5.74^{+0.15}_{-0.24}$ & $7.46\pm0.02$\\ 
UM133 & 0.0053 & $7.46^{+0.16}_{-0.25}$ & $-1.10^{+0.25}_{-0.69}$ & $7.74^{+0.15}_{-0.24}$ & $7.82\pm0.01$\\ 
UM311 & 0.0057 & $9.37^{+0.12}_{-0.16}$ & $-0.14^{+0.26}_{-0.73}$ & $8.90^{+0.15}_{-0.23}$ & $8.36\pm0.01$\\ 
UM448 & 0.0184 & $9.68^{+0.06}_{-0.07}$ & $0.86^{+0.27}_{-0.81}$ & $10.11^{+0.15}_{-0.23}$ & $8.32\pm0.01$\\ 
UM461 & 0.0034 & $7.08^{+0.15}_{-0.24}$ & $-1.82^{+0.25}_{-0.66}$ & $6.85^{+0.15}_{-0.24}$ & $7.73\pm0.01$\\ 
VIIZw403 & -0.0002 & $4.61^{+0.05}_{-0.06}$ & $-2.85^{+0.25}_{-0.62}$ & $5.62^{+0.15}_{-0.23}$ & $7.66\pm0.01$\\ 
\hline
\end{tabular}
\end{spacing}
\tablefoot{Column description: (1) source name; (2) redshift based on [CII] emission line; (3) stellar mass from CIGALE; (4) SFR from [CII] emission; (5) mass of total molecular gas estimated from [CII] luminosity; (6) metallicity values in $\mathrm{12+log(O/H)}$ from \cite{Madden13}. SFRs and $M_\mathrm{H2,TOT}$ were obtained from the [CII] luminosity of the narrow components following the relations found by \cite{DeLooze14} and \cite{Madden20} for the DGS galaxies, respectively. The typical $\mathrm{z_{[CII]}}$ uncertainties based on the spectral resolution are 0.0001.}
\end{center}
\end{table*}

\begin{table*}[h!]
%\begin{landscape}
%\begin{longtable}
\caption{Information on [CII] spectra acquisition and Gaussian fit.}
\label{tab:technical}
\begin{center}
\begin{spacing}{1.25}
\begin{tabular}{p{0.11\textwidth} c c c c c c c c c}
\hline
\hline
\multicolumn{9}{c}{Individual galaxies}\\\\
Source & OBSID & Extraction & $\mathrm{FWHM_{narrow}}$ & $\mathrm{FWHM_{broad}}$ & $\mathrm{log(L_{[CII],narrow})}$ & $\mathrm{log(L_{[CII],broad})}$ & $\chi^2_{single}$ & $\chi^2_{double}$\\
 & & & [km~s$^{-1}$] & [km~s$^{-1}$] & [L$_{\odot}$] & [L$_{\odot}$] & &\\
 (1) & (2) & (3) & (4) & (5) & (6) & (7) & (8) & (9)\\
\hline
Haro2 & 1342230081 & $5\times5$ & $255\pm2$ & $965\pm193$ & $7.10^{+0.01}_{-0.01}$ & $6.06^{+0.09}_{-0.12}$ & 1.21 & 0.68\\
Haro3 (*) & 1342221892 & $5\times5$ & $223\pm1$ & $306\pm61$ & $6.74^{+0.01}_{-0.01}$ & $6.15^{+0.17}_{-0.29}$ & 1.49 & 1.10\\
Haro11 & 1342199236 & $3\times3$ & $172\pm7$ & $570\pm56$ & $8.21^{+0.01}_{-0.01}$ & $7.53^{+0.09}_{-0.12}$ & 3.59 & 1.19\\
He2-10 & 1342221975 & $5\times5$ & $244\pm2$ & $580\pm33$ & $7.26^{+0.01}_{-0.01}$ & $6.60^{+0.05}_{-0.06}$ & 9.60 & 0.99\\
HS0052+2536 & 1342213134 & central & $211\pm23$ & - & $7.14^{+0.06}_{-0.07}$ & - & 0.87 & -\\
HS1222+3741 & 1342232306 & central & $147\pm32$ & $159\pm83$ & $6.33^{+0.12}_{-0.17}$ & $6.13^{+0.22}_{-0.49}$ & 1.04 & 0.97\\ 
HS1304+3529 & 1342199736 & $3\times3$ & $191\pm17$ & - & $6.55^{+0.05}_{-0.06}$ & - & 0.92 & -\\ 
HS1330+3651 & 1342199734 & $3\times3$ & $239\pm28$ & - & $6.76^{+0.03}_{-0.03}$ & - & 1.17 & -\\ 
HS1442+4250 & 1342208927 & $3\times3$ & $281\pm59$ & - & $4.38^{+0.10}_{-0.13}$ & - & 0.79 & -\\ 
HS2352+2733 & 1342213133 & central & $270\pm50$ & - & $6.11^{+0.09}_{-0.12}$ & - & 0.84 & -\\ 
IZw18  & 1342220973 & $3\times3$ & $200\pm23$ & - & $4.56^{+0.06}_{-0.07}$ & - & 0.97 & -\\ 
IIZw40 & 1342228253 & $5\times5$ & $236\pm1$ & - & $6.53^{+0.01}_{-0.01}$ & - & 0.87 & \\ 
Mrk153  & 1342209015 & $3\times3$ & $257\pm5$ & - & $6.34^{+0.01}_{-0.01}$ & - & 0.89 & -\\ 
Mrk209  & 1342199423 & $3\times3$ & $214\pm6$ & - & $4.57^{+0.02}_{-0.02}$ & - & 0.92 & -\\ 
Mrk930  & 1342212520 & $3\times3$ & $253\pm9$ & $466\pm201$ & $7.63^{+0.03}_{-0.03}$ & $6.78^{+0.30}_{-1.91}$ & 0.90 & 0.87\\
Mrk1089 (*) & 1342217859 & $5\times5$ & $238\pm1$ & $515\pm58$ & $7.94^{+0.01}_{-0.01}$ & $7.33^{+0.09}_{-0.12}$ & 2.71 & 1.05\\
Mrk1450 & 1342222070 & $3\times3$ & $245\pm14$ & - & $5.59^{+0.03}_{-0.04}$ & - & 0.88 & -\\ 
SBS0335-052 & 1342214221 & central & $160\pm26$ & - & $5.64^{+0.08}_{-0.10}$ & - & 1.04 & -\\ 
SBS1159+545 & 1342199228 & central & $231\pm30$ & - & $5.59^{+0.07}_{-0.08}$ & - & 0.82 & -\\ 
SBS1211+540 & 1342199422 & central & $249\pm69$ & - & $4.54^{+0.08}_{-0.09}$ & - & 0.78 & -\\ 
SBS1249+493 & 1342232266 & $3\times3$ & $244\pm29$ & - & $6.34^{+0.07}_{-0.08}$ & - & 0.94 & -\\ 
SBS1415+437 & 1342199733 & $3\times3$ & $247\pm11$ & - & $5.03^{+0.03}_{-0.03}$ & - & 0.87 & -\\ 
SBS1533+574 & 1342199230 & $3\times3$ & $207\pm6$ & - & $6.57^{+0.02}_{-0.02}$ & - & 0.90 & -\\ 
UGC4483 & 1342203684 & $3\times3$ & $189\pm12$ & $279\pm117$ & $3.73^{+0.03}_{-0.04}$ & $2.92^{+0.25}_{-0.62}$ & 1.05 & 1.04\\
UM133 & 1342212533 & $3\times3$ & $177\pm33$ & - & $5.79^{+0.04}_{-0.04}$ & - & 0.78 & -\\ 
UM311 & 1342213288 & $5\times5$ & $241\pm4$ & $369\pm58$ & $6.99^{+0.02}_{-0.02}$ & $6.14^{+0.15}_{-0.22}$ & 1.08 & 0.86\\ 
UM448 (*) & 1342222201 & $3\times3$ & $237\pm5$ & $907\pm116$ & $8.24^{+0.01}_{-0.01}$ & $7.42^{+0.09}_{-0.12}$ & 7.51 & 0.92\\ 
UM461 & 1342222205 & central & $144\pm11$ & $672\pm355$ & $4.88^{+0.04}_{-0.05}$ & $4.64^{+0.17}_{-0.29}$ & 0.93 & 0.83\\ 
VIIZw403 & 1342199289 & $5\times5$ & $234\pm6$ & - & $3.60^{+0.01}_{-0.01}$ & - & 1.07 & -\\ 
\hline
\multicolumn{9}{c}{Stacking}\\\\
Whole sample & - & - & $247\pm5$ & $437\pm51$ & $6.45^{+0.02}_{-0.02}$ & $5.93^{+0.10}_{-0.14}$ & 1.91 & 0.81\\
Non-det. & - & - & $240\pm6$ & $478\pm87$ & $6.09^{+0.02}_{-0.02}$ & $5.48^{+0.14}_{-0.21}$ & 1.12 & 0.76\\
\hline
\end{tabular}
\end{spacing}
\tablefoot{Column description: (1) source name; (2) PACS Observation Identification number; (3) extraction region of the [CII] spectrum, i.e., from central spaxel, $3\times3$ or $5\times5$ spaxel coverage; (4) FWHM of the narrow component; (5) FWHM of the broad component; (6) [CII] luminosity of the narrow component; (7) [CII] luminosity of the broad component; (8) $\chi^2$ from single Gaussian fit; (9) $\chi^2$ from double Gaussian fit. Sources without $\mathrm{FWHM_{broad}}$, $\mathrm{L_{[CII],broad}}$, and $\chi^2_{double}$ do not show any broad component in their spectra. For such cases, $\mathrm{FWHM_{narrow}}$ and $\mathrm{L_{[CII],narrow}}$ correspond to the value obtained with a single Gaussian component. The [CII] spectrum of targets with (*) was modeled with a three-component Gaussian (see Sect. \ref{subsec:individual_out}). For these sources, $\mathrm{FWHM_{narrow}}$ and $\mathrm{L_{[CII],narrow}}$ are those obtained from a single Gaussian given by the sum of the two narrow components of the fit. In the bottom part of the table, first column refers to the sample used for stacking.}
\end{center}
%\end{longtable}
%\end{landscape}
\end{table*}

The final stellar masses and SFRs are reported in Table \ref{tab:summary} and shown in Fig. \ref{fig:ms} in the SFR-$\mathrm{M_{*}}$ diagram (e.g., \citealt{Noeske07,Rodighiero11,Speagle14}). For comparison, we show (as density contours) the sample of $\sim15,750$ nearby galaxies from \cite{Leroy19}, observed by the Wide-field Infrared Survey Explorer (WISE; \citealt{Wright10}) and by the Galaxy Evolution Explorer (GALEX; \citealt{Martin05}). Their best fit to the local star-forming main-sequence\footnote{The best-fit relation to the star-forming main-sequence in \cite{Leroy19} was obtained by selecting SFGs with $\mathrm{log(M_{*}/M_{\odot})=9.5-11}$. We show in Fig. \ref{fig:ms} their original fit, extrapolating the relation to the lower stellar mass regime probed by our dwarf galaxies.} and associated $3\sigma$ scatter is also reported. Most of our sources lie within and in the upper side of the main-sequence, with a few galaxies with larger stellar mass and SFR tending to the starburst-dominated region.

\section{Observations and data reduction}\label{sec:data}
We downloaded the fully-calibrated science-ready data of the 35 compact objects in our sample from the \textit{Herschel} Science Archive\footnote{\url{http://archives.esac.esa.int/hsa/whsa/}.}. In this study, we focused on [CII]~$158~\mathrm{\mu m}$ emission which mostly traces the atomic gas in our galaxies and was observed for the full sample. %However, we also gathered [OIII]~$88~\mathrm{\mu m}$ data which are available for 30 out of 29 sources, in order to constrain the ionized gas phase as well.

The field of view (FoV) of the PACS spectrometer is a footprint of $5\times5$ spatial pixels (spaxels) $9.4''$-sized, for a total coverage of $47'' \times 47''$. %The spectral resolution is $125~\mathrm{km~s^{-1}}$ and $240~\mathrm{km~s^{-1}}$ at $88~\mathrm{\mu m}$ and $158~\mathrm{\mu m}$, respectively. 
The spectral resolution is $240~\mathrm{km~s^{-1}}$ at $158~\mathrm{\mu m}$ which is consistent with the average full width at half maximum (FWHM) of our sample, allowing us to identify possible flux excess at higher velocities (see Sect. \ref{subsec:individual_out}). To obtain the spectra and extract the fluxes needed for our analysis, we used the \textit{Herschel} Interactive Processing Environment (HIPE; \citealt{Ott10}), version 15.0.1, on the re-binned cubes. These are the main science-use cubes produced by the HIPE standard pipelines, with the native footprint of the PACS FoV and an increased spectral resolution as defined by the parameters \textit{upsample} and \textit{oversample}. %By default, oversample and upsample are set to 2 and 4, respectively, resulting in a final sampling in the spectral direction of $16~\mathrm{km~s^{-1}}$ at $88~\mathrm{\mu m}$ and $30~\mathrm{km~s^{-1}}$ at $158~\mathrm{\mu m}$ (see PACS documentation for more). 
By default, these two quantities are set to 4 and 2, respectively, resulting in a final sampling in the spectral direction of $30~\mathrm{km~s^{-1}}$ at $158~\mathrm{\mu m}$ (see PACS documentation for more). The re-binned cubes are unique for pointed observations of the most compact objects. For slightly more extended galaxies, mapping observations were made and multiple cubes are created for each source. In those cases, we plotted the different footprints on the corresponding PACS photometry image, taking only the re-binned cube in which the source was located at (on near to) the central spaxel.

We obtained three different spectra from each data cube with the HIPE tool. In particular, we used standard HIPE tasks to extract the spectrum from \textit{i}) the central spaxel, \textit{ii}) the sum of the $3\times3$ inner spaxels, and \textit{iii}) the sum of the full $5\times5$ spaxel coverage. %To each spectrum, a point-source correction is applied to take into account the size of the PACS beam (i.e., $\mathrm{FWHM\sim9.5''}$ at $88~\mathrm{\mu m}$, and $\sim12''$ at $158~\mathrm{\mu m}$) and the flux loss between spaxels (due to the fact that the PACS footprint is not regular and the spaxels are not contiguous to each other). 
To each spectrum, a point-source correction was applied to take into account the size of the PACS beam (i.e., $\sim12''$ at $158~\mathrm{\mu m}$) and the flux loss between spaxels (due to the fact that the PACS footprint is not regular and the spaxels are not contiguous to each other). By looking at the spatial extension of each source across the PACS footprint and by comparing the three spectra, we took the one with the largest flux and a high signal-to-noise ($S/N$) in order to include as much information as possible from the line emission. For 6 out of 35 sources, the final spectrum was too noisy to be characterized, thus we excluded such objects from our analysis (see Appendix \ref{app:bad_spectra}). Therefore, the final sample is composed of 29 galaxies, whose properties are reported in Tables \ref{tab:summary} and \ref{tab:technical}. We made a first Gaussian fit of the line to have an initial estimate of its FWHM, and used this value to define the continuum emission as the region between $2-5\times FWHM$ (both redward and blueward of the line peak), so to avoid the possible wings of the outflow component and the noisy end of the spectrum. To fit the emission profiles we used \texttt{SCIPY.OPTIMIZE.CURVE\_FIT} \citep{Virtanen20}, providing as initial guesses for the center, peak, and FWHM of the line, the expected wavelength of [CII] emission (based on the available spectroscopic redshift of each source), the maximum flux value around that position, and the PACS spectral resolution. Because the shape of the continuum emission could be rather variable in our spectra, we followed \citealt{Lebouteiller12,Cormier15} and modeled the continuum with a polynomial curve of order 1 or 2, or with a Chebyshev series\footnote{The Chebyshev series is defined as $P(x)= \sum_{i=0}^{i=3} C_i \cdot T_i(x)$, where $T_i(x)$ is the Chebyshev polynomial of first kind and degree $i$, and $C_i$ are the coefficients of the fitting.} of third degree \citep{Rivlin74}. Then, we simultaneously fit the line and continuum for each source, taking as the best continuum modeling the one providing the minimum reduced $\chi^2$, after further checking the goodness of the fit through visual inspection. Finally, we subtracted the continuum to each spectrum in order to better analyze the line and wings profiles.

\section{Analysis and results}\label{sec:analysis}
We started searching for outflow signatures in the spectra of each individual galaxy by fitting their [CII] emission lines with a single and double Gaussian profile (the latter including a narrow and broad component), and by inspecting the corresponding residuals. In this case, we use the output value of the central wavelength obtained from the fit in the previous section as initial guess for both the Gaussian components, leaving the peak flux as a free parameter. We then adopt the PACS resolution as guess for the FWHM of the narrow component, and twice its value for the broad component. In some cases, the presence of outflowing gas is clearly evidenced by the broad component which, by matching the high-velocity wings of the spectra, improves the quality of the fit. To quantify such an improvement, we compared the reduced $\chi^2$ of the fit with the single ($\chi^2_\mathrm{single}$) and double ($\chi^2_\mathrm{double}$) Gaussian profiles, considering the presence of the possible outflow component only when $\chi^2_\mathrm{double}< \chi^2_\mathrm{single}$. As a result, we found that 11 out of 29 galaxies show clear signs of outflowing atomic gas as traced by the [CII] emission, while in the remaining sources the outflow (if present) is too faint to be individually detected (see Sect. \ref{subsec:spec_stacking}). In the following, we first compute the properties of the 11 sources with outflow evidence, and then we estimate the average outflow features from the entire galaxy sample through line and cube stacking.

\subsection{Individual outflow detection}\label{subsec:individual_out}
The galaxies with clear outflow detections are characterized by evident wings in the spectra at typical velocities of $\pm 400~\mathrm{km~s^{-1}}$, where the residuals between the single Gaussian model and the line present an excess of emission. In these cases, such residuals are reduced to the noise level by adopting a double Gaussian profile, including a broad component to account for the high-velocity wings in the spectra (see Appendix \ref{app:individual_fit}).

We retrieved different quantities from the [CII] spectra that we use in the following sections to investigate the strength and efficiency of the outflow in the 11 galaxies described above. First, we computed the FWHM of the narrow and broad component as $FWHM=(FWHM_{obs}^2 - FWHM_{inst}^2)^{1/2}$, where $FWHM_{obs}=2.355\sigma$ is the observed width of the line (with $\sigma$ the standard deviation of the corresponding Gaussian function), and $FWHM_{inst}$ is the PACS instrumental line width (i.e., $240~\mathrm{km~s^{-1}}$ for [CII])\footnote{We applied the deconvolution for the spectral resolution of the instrument only to intrinsic line widths larger than 150~km~s$^{-1}$ (see \citealt{Cormier15}).}. We found that the broad component has, on average, a FWHM more than two times larger than the narrow component, with means of $\sim526$ and $213~\mathrm{km~s^{-1}}$, respectively. Then, we obtained the [CII] luminosity of both components by following \cite{Solomon92} as:
\begin{equation}
    L_\mathrm{{[CII]}} = 1.04 \times 10^{-3}~S_\mathrm{[CII]}~\Delta v~D_L(z_\mathrm{[CII]})^2~\nu_{obs}~[\mathrm{L_\odot}],
    \label{eq:line_lum}
\end{equation}
where $S_\mathrm{[CII]}~\Delta v$ is the velocity-integrated line flux in units of Jy~km~s$^{-1}$, $\nu_{obs}$ is the observed peak frequency in GHz, and $D_L$ is the luminosity distance in Mpc at the redshift derived from the centroid of the single Gaussian fit of the [CII] line (i.e., $z_\mathrm{[CII]}$). In particular, we used the [CII] luminosity of the narrow component to estimate the SFR$_\mathrm{[CII]}$ of each galaxy (see Sect. \ref{sec:sample}). To obtain the error on $\mathrm{L_{[CII]}}$, we perturbed the [CII] integrated flux $\mathrm{N}=1000$ times within its uncertainty. We then took the 16th and 84th percentiles of the resulting distribution as the error on $S_\mathrm{[CII]}~\Delta v$, propagating it to $\mathrm{L_{[CII]}}$ based on Eq. \ref{eq:line_lum}.

We found that, for three galaxies (i.e., Haro3, Mrk1089, and UM448), a three-Gaussian fit (with two narrow and one broad components) provided better residuals than obtained from the two-Gaussian fit, improving the modeling of their global [CII] profiles. Interestingly, the morphology and kinematics of these sources show evidence for merging activity (e.g., \citealt{Dopita02,Johnson00,Johnson04,Cairos07,Galametz09,James13}). For instance, \cite{James13} made use of the Fibre Large Array Multi Element Spectrograph (FLAMES; \citealt{Pasquini02}) at the Very Large Telescope (VLT) to detect H$\alpha$ emission over UM448. They found a complex emission line profile, composed of two narrow components and a third broad component possibly associated with an outflow. Similarly, we modeled the [CII] spectrum of UM448 adding a third narrow Gaussian component to the fit. As the goodness of this fit was better than the double-Gaussian one for all the three sources and given the evidence of mergers, for the rest of the paper we use the parameters from the three-component modeling for Haro3, Mrk1089, and UM448. A few examples of individual spectral fitting for the galaxies in our sample are shown in Appendix \ref{app:individual_fit}. All the quantities derived in this section are reported in Table \ref{tab:technical}.

\begin{figure}
    \begin{center}
	\includegraphics[width=\columnwidth]{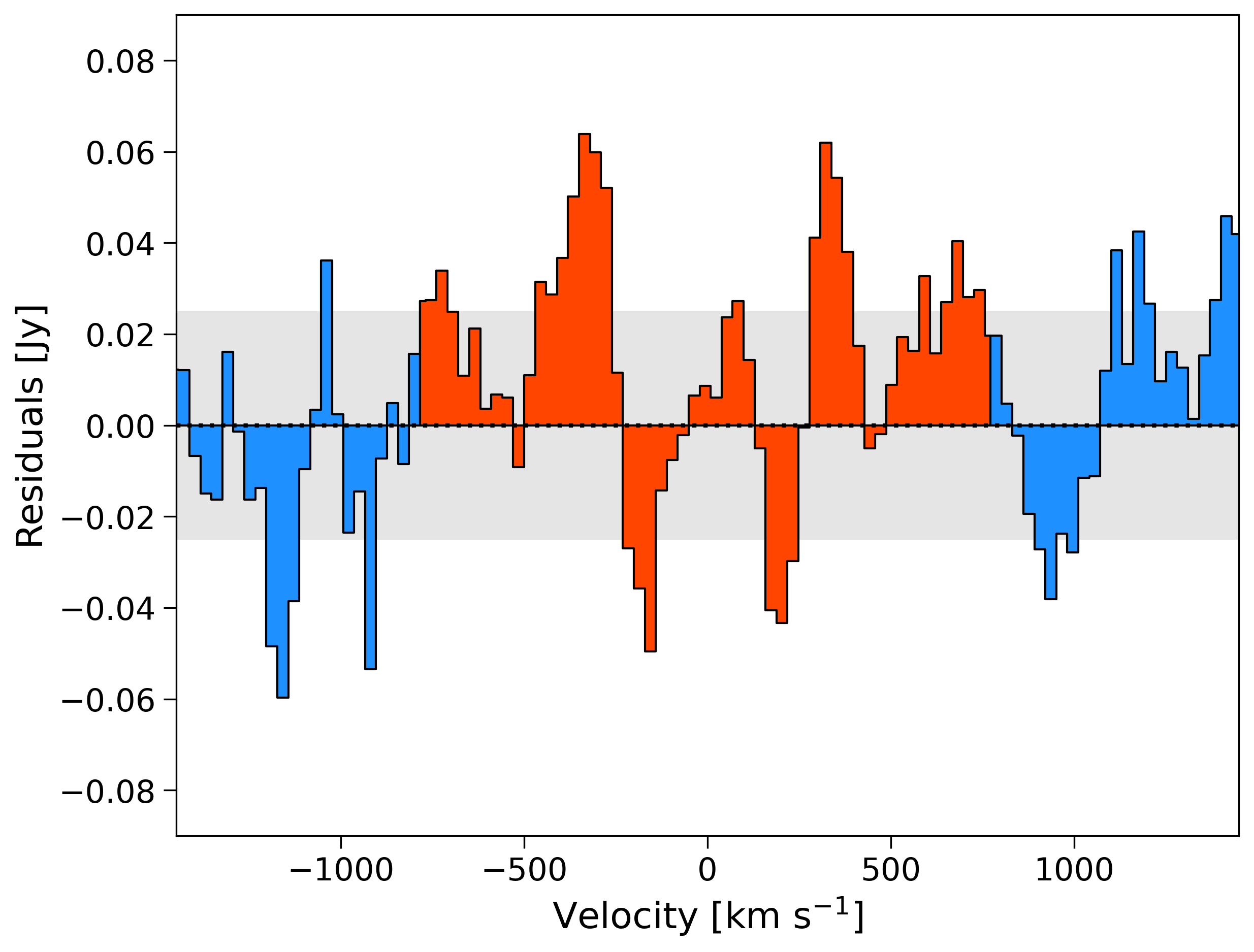}
	\end{center}
    \caption{Variance-weighted stacked residuals obtained after subtracting the best-fit single component Gaussian function to each [CII] spectrum. Red bins represent the residuals within $\pm800~\mathrm{km~s^{-1}}$, whereas the blue bins mark the spectral region used for estimating the noise. The dotted horizontal line sets the zero level, while the shaded area represents the noise at $\pm1\sigma$. A $\sim3\sigma$ excess is visible at velocities $\pm400~\mathrm{km~s^{-1}}$, suggesting that a further Gaussian component is needed to model the [CII] emission in our galaxies.}
    \label{fig:stacked_res}
\end{figure}
\subsection{Spectral stacking}\label{subsec:spec_stacking}
\begin{figure*}[t]
    \begin{subfigure}{0.5\textwidth}
    \centering
	\includegraphics[width=0.9\linewidth]{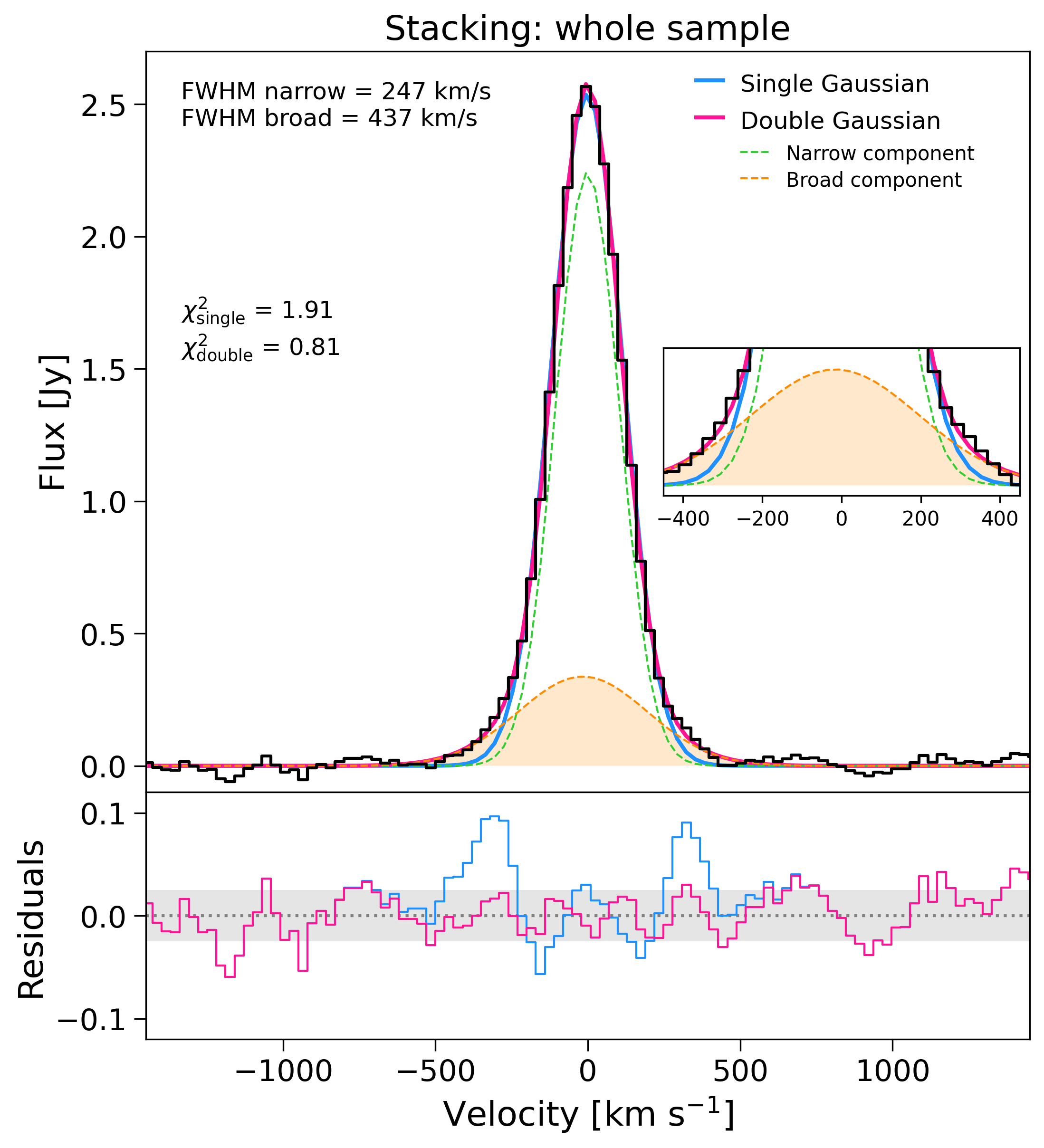}
	\end{subfigure}%
	\begin{subfigure}{0.5\textwidth}
	\centering
	\includegraphics[width=0.9\linewidth]{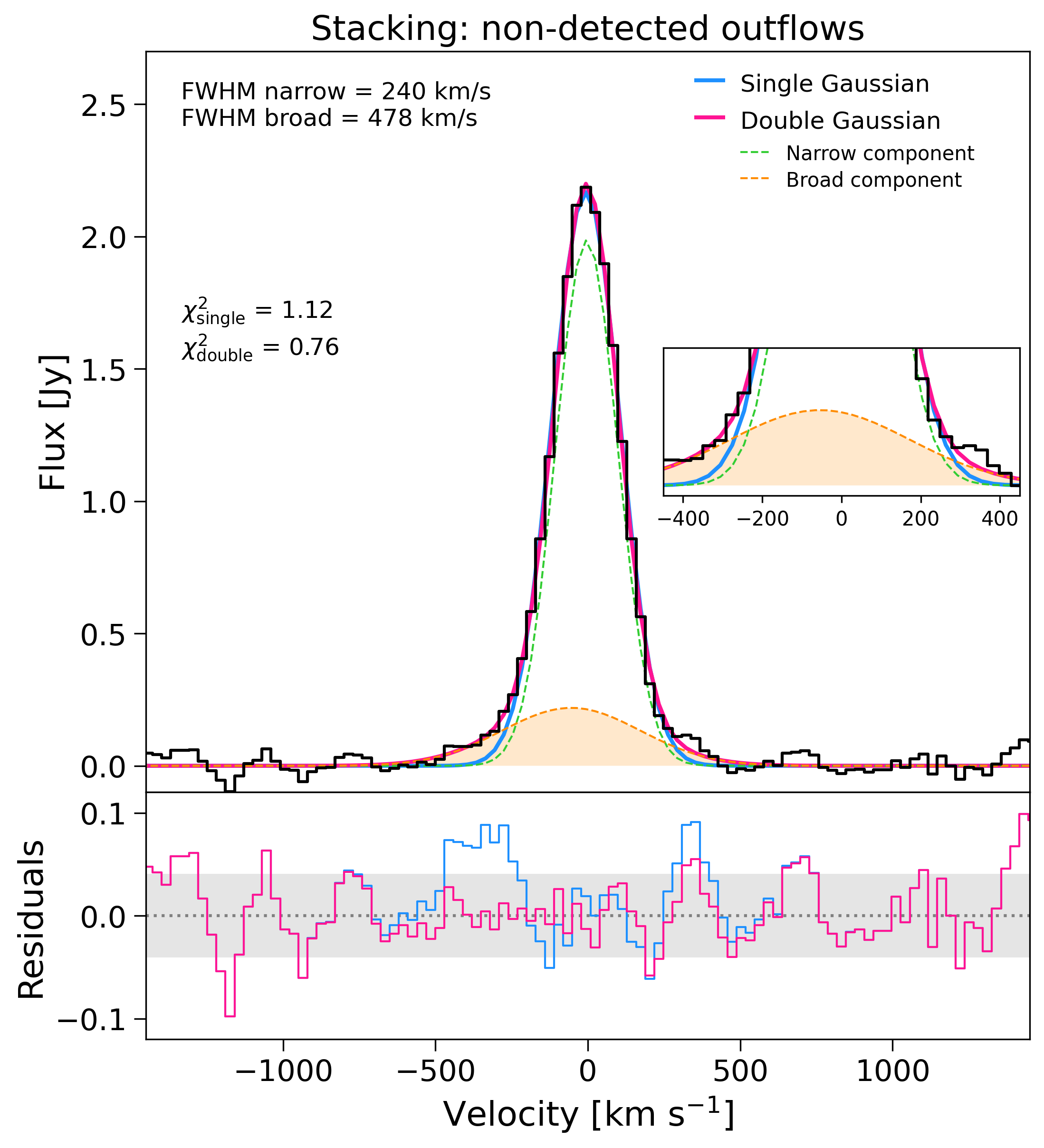}
	\end{subfigure}
	\caption{\textit{Left:} Stacked variance-weighted [CII] spectrum (black histogram) of the whole sample as a function of velocity. Both the fit with a single Gaussian function (in blue) and with a double Gaussian profile (in pink) are reported. The latter is the sum of a narrow (green line) and broad (orange line and shaded area) component. The FWHM of both components and the corresponding reduced $\chi^2$ are also shown in the figure. A zoom-in of the spectral region dominated by outflows is shown as an inset plot on the right. The bottom panel reports the residuals from the single (blue) and double (pink) Gaussian functions. The dotted horizontal line marks the zero level, while the shaded area represents the noise of each spectrum at $\pm1\sigma$, computed as described in the text. \textit{Right:} same as left panel, but for the stacking of only the sources with non-detected outflows.}\label{fig:spec_stacking}
\end{figure*}
In order to get the average outflow properties of our full galaxy sample and to investigate the possible presence of the outflows in the 18 sources with no individual detection, we performed a stacking of their [CII] spectra. First, we used the computed $z_\mathrm{[CII]}$ to align each continuum-subtracted spectrum to the [CII] rest-frame frequency. Then, following e.g., \citealt{Gallerani18,Ginolfi20}, we tested the null hypothesis that the [CII] line profiles in our sample can be fully reproduced by a single Gaussian component. To do so, we computed the residuals of each galaxy by subtracting the best-fitting Gaussian function to the corresponding observed flux. We thus combined the obtained residuals with a variance-weighted stacking as
\begin{equation}
    R_{\mathrm{stack}} = \frac{\sum_{i=1}^{N} R_i \cdot w_i}{\sum_{i=1}^{N} w_i},
    \label{eq:res_stacking}
\end{equation}
where $R_i$ represents the residual of the $i$-th galaxy, $N$ is the number of stacked sources, and $w_i=1/\sigma^2_{i}$ is the weighting factor (with $\sigma_i$ the noise associated to each spectrum). To estimate $\sigma_i$, we avoided the velocity range [-800; +800]~$\mathrm{km~s^{-1}}$ in order to exclude contamination from the [CII] emission and the broad wings. 

\begin{figure*}
    \centering
    \begin{subfigure}[b]{0.8\textwidth}
	\includegraphics[width=1\linewidth]{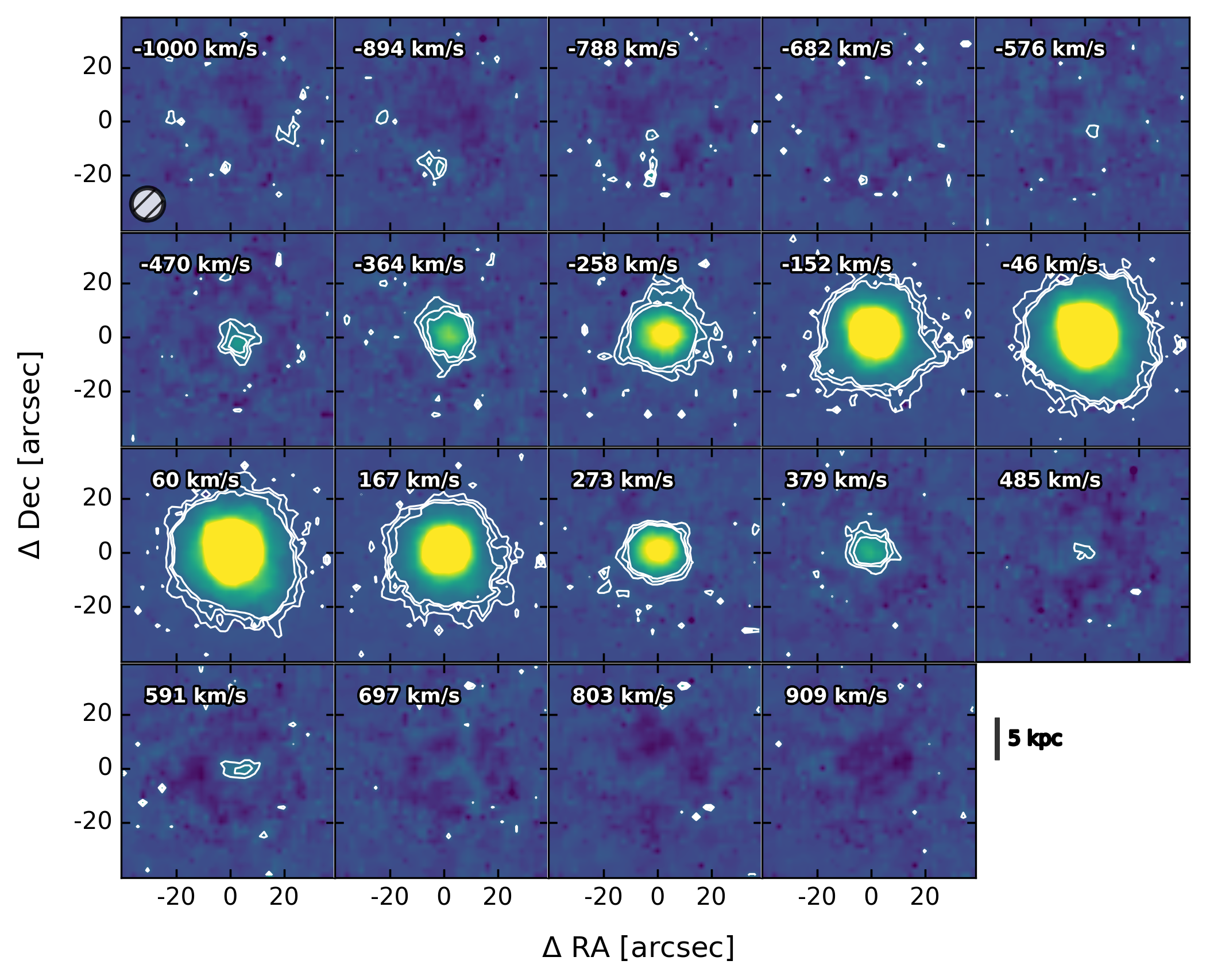}
    \end{subfigure}
    
    \begin{subfigure}[b]{0.75\textwidth}
	\includegraphics[width=1\linewidth]{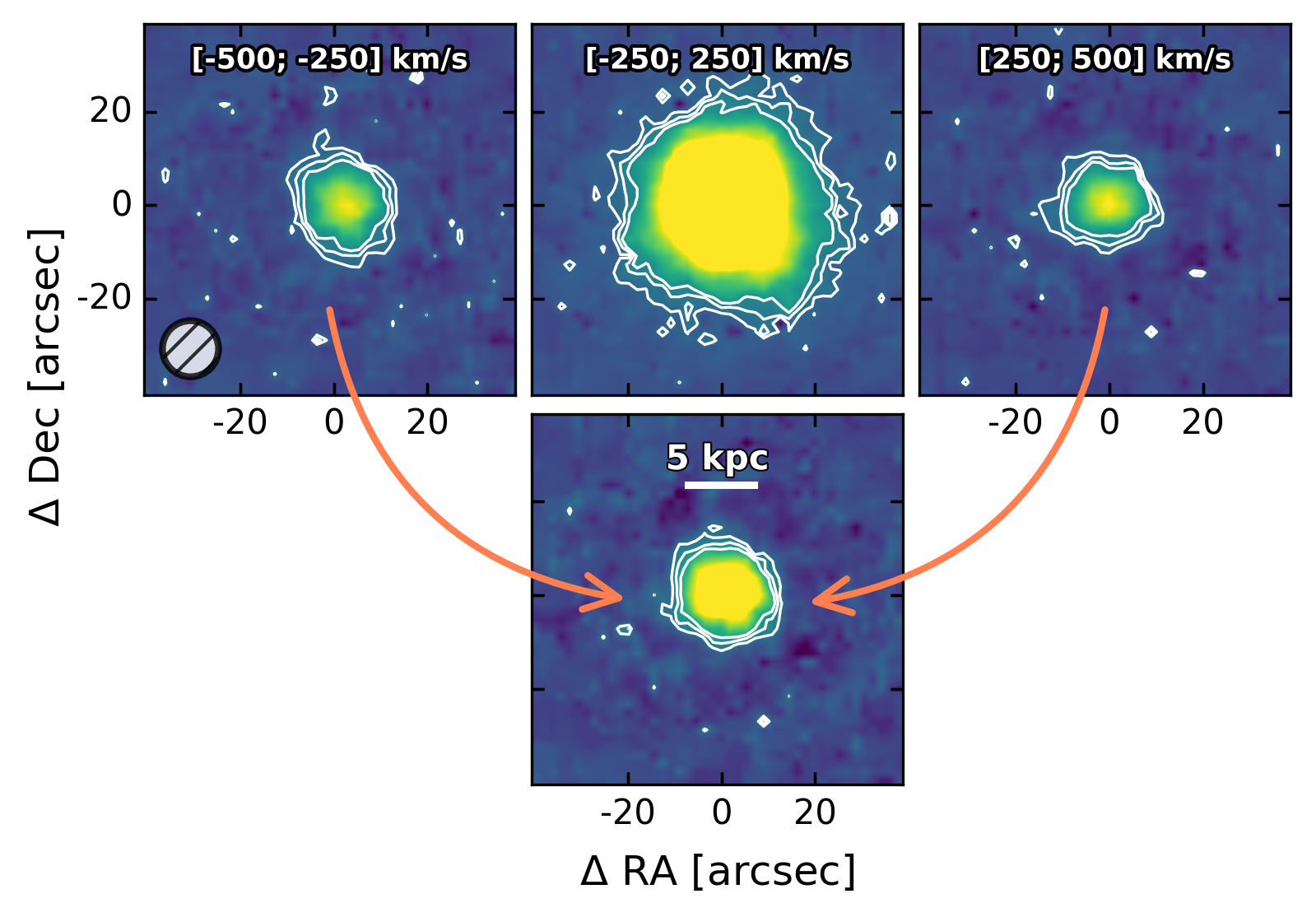}
    \end{subfigure}
\caption{\textit{Top:} Channel maps of the stacked cube covering $\sim2000~$km~s$^{-1}$ around the peak of the emission line. Velocity bins are in steps of $\sim106~$km~s$^{-1}$ for a better representation. Each spectral channel shows the [CII] emission from a $80''\times80''$ region. Contour levels are shown in white at 3, 5, and 7 $\sigma$, where $\sigma$ is the rms computed in each channel. \textit{Bottom:} [CII] integrated intensity maps of the outflow and core emission. Left and right panels are obtained by summing the emission of the broad wings in the velocity ranges [-500, -250] and [250; 500]~km~s$^{-1}$, respectively, while the central panel represents the core emission at [-250; 250]~km~s$^{-1}$. The bottom panel is the sum of the two velocity-integrated maps of the broad wings (as pointed by the arrows), representing the whole outflow emission. Contour levels are shown in white at 3, 5, and 7 $\sigma$, where $\sigma$ is the rms of the integrated intensity map. Both figures report the PACS beam (as shown in the lower-left corner of the first panel), and a reference scale of 5~kpc.}
\label{fig:channel_maps}
\end{figure*}

The resulting stacked residuals are shown in Fig. \ref{fig:stacked_res}, where a clear excess of emission is evidenced by the two peaks at velocities $\pm400~\mathrm{km~s^{-1}}$, reaching a significance of $\sim3\sigma$. This proves that an additional component is needed in order to reproduce the observed [CII] fluxes. Indeed, if our spectra were fully characterized by a single Gaussian function, the residuals in Fig. \ref{fig:stacked_res} should be consistent with the noise over the full spectral range (see left panel of Fig. \ref{fig:fitting_example}). Two weaker negative peaks are also visible at velocities $\pm200~\mathrm{km~s^{-1}}$. As also discussed by \cite{Ginolfi20}, these represent another signature of the poor single Gaussian fits to the [CII] line profiles of our galaxies, which tend to underestimate the flux at low velocities in order to attribute some flux at the high-velocity wings (see e.g., middle and right panels in Fig. \ref{fig:fitting_example}).

At this point, we proceeded with a stacking of the [CII] spectra of all the galaxies in our sample. Similarly to what done for the residuals, we defined the stacked spectrum as
\begin{equation}
    S_{\mathrm{stack}} = \frac{\sum_{i=1}^{N} S_i \cdot w_i}{\sum_{i=1}^{N} w_i},
    \label{eq:line_stacking}
\end{equation}
where $S_i$ is the flux of the $i$-th galaxy, and all the other parameters are the same as for Eq. \ref{eq:res_stacking}. Fig. \ref{fig:spec_stacking} (left panel) shows the result of this procedure. As done for the individual outflow detections, we fitted the stacked spectrum with both a single and double Gaussian profile, comparing the corresponding reduced $\chi^2$ and residuals. The spectrum shows clear signs of broad wings at velocities $\pm\sim400$~km~s$^{-1}$, as evidenced by the corresponding large residuals obtained by using a single Gaussian function to fit the line profile. The two-component model clearly improves the fit, resulting in a better reduced $\chi^2$ and a residual flux consistent with the noise over the entire velocity range.

To ensure that the stacking result was not biased by the presence of a few sources with stronger evidence of outflows, we performed a delete-$d$ jackknife resampling \citep{Shao89}. We recomputed 500 times the stacked spectrum by excluding each time $\sim10\%$ of the sample (i.e., 3 galaxies), in order to get an estimate of the wings variation while still preserving a large enough sample to stack. The resulting FWHM distributions of both Gaussian components are in agreement with what obtained by stacking the whole sample, implying that our results are not affected by outliers. We repeated the stacking by including only the 18 sources with no individual outflow detection to investigate the presence of broad wings in this sub-sample. This is shown in Fig. \ref{fig:spec_stacking} (right panel), where the high-velocity tails are still visible in the spectrum and recovered with a broad component, although they are weaker than those found in the stack of the whole sample. Again, the jackknife statistics did not find any track of outliers, as expected given that none of the galaxies in this sub-sample show significant evidence of outflowing gas.

It is interesting to note that most of the galaxies with individual outflow detections lie at the top-right corner of the main-sequence diagram, with the largest stellar masses and SFRs (see Fig. \ref{fig:ms}). This suggests that the stronger broad component in the stacked spectrum of the whole sample could be driven by sources with the largest star-formation activity, as also found at high redshift \citep{Ginolfi20} and as expected by the well-known [CII]-SFR relation (e.g., \citealt{DeLooze14,Schaerer20,Romano22}).

As done for the individual sources showing broad wings, we estimated the [CII] luminosity of the broad component of both the stacked spectra through Eq. \ref{eq:line_lum}, as well as their FWHM. These values are listed in Table \ref{tab:technical} and they will be used later to constrain the outflow efficiency of the average population of dwarf galaxies.

%\begin{figure}
%    \begin{center}
%	\includegraphics[width=\columnwidth]{plots/stacking_invweights_new.png}
%	\end{center}
%    \caption{Stacked variance-weighted [CII] spectrum (black histogram) of the whole sample as a function of velocity. Both the fit with a single Gaussian function (in blue) and with a double Gaussian profile (in pink) are reported. The latter is the sum of a narrow (green line) and broad (orange line) component. The FWHM of both components and the corresponding reduced $\chi^2$ are also shown in the figure. The bottom panel report the residuals from the single (blue) and double (pink) Gaussian functions. The dotted horizontal line marks the zero level, while the shaded area represents the noise of each spectrum at $\pm1\sigma$, computed as described in the text. \textbf{[FWHM is not corrected for intrinsic res]}}
%    \label{fig:spec_stacking}
%\end{figure}

\subsection{Spatial stacking}\label{subsec:cube_stacking}
To characterize the average spatial extent of the atomic outflows in our galaxies, we produced a stacked [CII] cube. As done for the spectra in Sect. \ref{subsec:spec_stacking}, we first aligned the spectral axes of each continuum-subtracted cube to the [CII] rest-frame emission. Then, we also spatially aligned the cubes by centering them on the peak of the corresponding [CII] intensity map produced by summing the fluxes from the spectral channels including the emission line. We used a variance-weighted stacking as in Eq. \ref{eq:line_stacking}, where now the $\sigma$ in the weighting factor represents the spatial rms estimated in each channel of the cube in regions free of emission. 

We show in Fig. \ref{fig:channel_maps} (top panel) the channel maps of the [CII] emission of the stacked cube from the central $80''\times 80''$ region, and within $\sim2000$~km~s$^{-1}$ of the spectral line. The [CII] emission is clearly detected at high velocities (i.e., where the broad wings in the [CII] stacked spectrum arise), with a bulk in the core of the line at $v\sim[-250;250]~$km~s$^{-1}$. As our aim is to obtain the average size of the outflow, we produced velocity-integrated [CII] maps of the wings in the low- and high-velocity tails at [-500, -250] and [250; 500]~km~s$^{-1}$, respectively (Fig. \ref{fig:channel_maps}, bottom panels), which resulted in high-significance detections of $\gtrsim20\sigma$. We then summed the two maps together to obtain the total outflow emission (detected at $\gtrsim30\sigma$), used to obtain the average outflow radius. In particular, we fitted a 2D Gaussian function to the total intensity map of the wings obtaining the outflow circularized effective radius defined as $R_{\mathrm{out}} = \sqrt{ab}$, where $a$ and $b$ are the best-fit beam-deconvolved semi-major and semi-minor axis of the Gaussian, respectively. We found $R_{\mathrm{out}}=0.99\pm0.18$~kpc, where the uncertainty was computed through the errors of $a$ and $b$ from the fit. Our result is in good agreement with previous assumptions and estimations from the literature in local galaxies (e.g., \citealt{Arribas14,Fluetsch19,Marasco22}).

The core of the stacked [CII] line emission results to be quite extended. By fitting the corresponding intensity map with the same method adopted for the estimation of the outflow radius, we obtained a deconvolved size of $R_{\mathrm{[CII]}}=1.49\pm0.05$~kpc, with some residuals surrounding the edge of the core suggesting the presence of a more extended emission. We thus compared the average [CII] size of our galaxies with the stellar distribution as traced by their rest-frame UV emission. To measure the individual UV sizes of our sources, we used the NUV band ($\lambda_\mathrm{mean}\sim2345~\AA$) photometry from GALEX, obtaining an average of $R_{\mathrm{UV}}=0.67\pm0.03$~kpc. Overall, we found that the [CII] emission in our dwarf galaxies is $\sim2$ times more extended then the UV. Interestingly, these results are in good agreement with those found for SFGs at $z\gtrsim4$ (e.g., \citealt{Fujimoto20,Herrera20,Lambert23}) which suggest the presence of circumgalactic [CII] halos likely produced by galactic outflows or past merging activity (e.g., \citealt{Fujimoto19,Fujimoto20,Ginolfi20}). We will further explore these results in a future work (Romano et al. in prep.).

%It is worth noting that the resulting stacked cube could combine outflows of different morphology, size, and orientation. However, it is still possible to get an average estimate of the radius of the outflow, posing a constraint on its spatial scale.

%%%%%%%%%%%%%%%%%%%%%%%%%%%%%%%%%%%%%%%%%%%%%%%%%%%%%%%%

\section{Discussion}\label{sec:discussion}

\subsection{Outflow efficiency}\label{subsec:outflow_efficiency}
\begin{figure*}
    \begin{center}
	\includegraphics[width=0.6\textwidth]{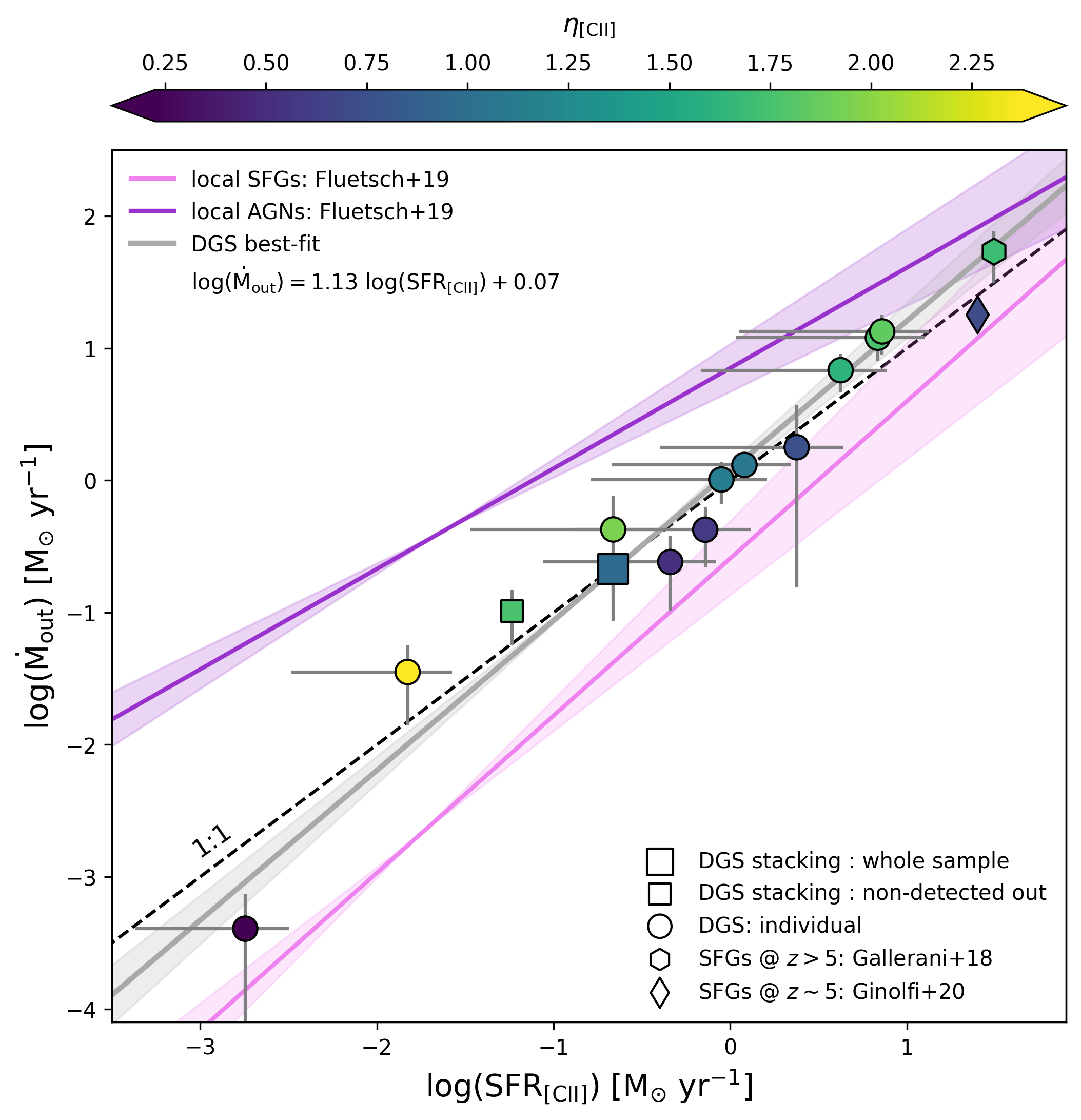}
	\end{center}
    \caption{Atomic outflow rate as a function of the SFR, for both individual detections of broad wings (circles) and from line stacking of the whole sample and of the sources with non-detected outflow (big and small squares, respectively). The pink and violet lines are the best-fit relations between molecular outflow rate and SFR for local AGN hosts and star-forming/starburst galaxies by \cite{Fluetsch19}, respectively, while the shaded regions are the corresponding uncertainties. The solid grey line with the shaded area represent a linear fit to the DGS galaxies with individual outflow detections and its uncertainty, respectively. The dashed line reports the 1:1 relation. We also show the results from [CII] stacking of $z\gtrsim5$ SFGs by \cite{Gallerani18} (hexagon) and \cite{Ginolfi20} (diamond). All markers are color-coded for their mass-loading factors.}
    \label{fig:outflow_sfr}
\end{figure*}

To fully characterize the outflows and their impact on the evolution of dwarf galaxies, a key parameter is the so-called mass-loading factor, i.e., the ratio between the rate of gas mass expelled out of the galaxy and the rate of star formation ($\eta=\dot{M}_\mathrm{out}$/SFR). This quantity is an estimate of the outflow efficiency and it represents a fundamental ingredient for simulations trying to explain the baryon cycle in galaxies.

We used the [CII] luminosity of the broad component (both for individual outflow detections and stacked spectra) to estimate the mass of the outflowing atomic gas (e.g., \citealt{Maiolino12,Bischetti19,Ginolfi20}). In particular, we considered the following relation by \cite{Hailey10}:
\begin{equation}
\begin{aligned}
    M_{\mathrm{out}}/M_{\odot} ={} & 0.77 \left(\frac{0.7 L_{\mathrm{[CII],broad}}}{L_{\odot}} \right) \left(\frac{1.4\times10^{-4}}{X_{\mathrm{C^{+}}}} \right)\\
    & \times \frac{1+2e^{-91K/T}+n_{\mathrm{crit}}/n}{2e^{-91K/T}},
    \end{aligned}
    \label{eq:outflow_mass}
\end{equation}
where $X_{\mathrm{C^{+}}}$ is the $\mathrm{C^{+}}$ abundance per hydrogen atom, $n_{\mathrm{crit}}\sim3\times10^{3}~$cm$^{-3}$ is the critical density of the [CII] transition (e.g., \citealt{Carilli13}), $T$ and $n$ are the gas temperature and density, respectively. Eq. \ref{eq:outflow_mass} was derived under the assumption of an optically thin [CII] emission (e.g., \citealt{Hailey10,Cicone15,Ginolfi20}), and assuming $X_{\mathrm{C^{+}}}=1.4\times10^{-4}$ \citep{Savage96}, $T$ in the range 60-200~K (see \citealt{Ginolfi20} and references therein), and $n\gg n_{\mathrm{crit}}$, all typical of PDRs. In addition, the factor 0.7 in the parenthesis of Eq. \ref{eq:outflow_mass} represents the fraction of [CII] emission arising from PDRs, while the remaining 30\% is supposed to come from the other phases of the ISM (e.g., \citealt{Stacey91,Stacey10,Diaz17,Cormier19}).

It is worth noting that the assumptions on the physical properties of the outflowing gas (i.e., the optically thin emission, the large number density) are conservative (as discussed in \citealt{Maiolino12}), providing us lower limits on $M_{\mathrm{out}}$. Furthermore, in this work we are only accounting for the atomic gas as traced by the [CII] emission, not considering that part of the outflow could be composed by the other ISM phases (i.e., molecular and ionized gas) that may contribute as well to the evolution of the host galaxy (e.g., \citealt{Veilleux05,Maiolino12,Muratov15,Fluetsch19}).

We thus computed the atomic mass outflow rate within the time-averaged expelled shells or clumps scenario \citep{Rupke05b}:
\begin{equation}
    \dot{M}_{\mathrm{out}} = \frac{v_{\mathrm{out}}\times M_{\mathrm{out}}}{R_{\mathrm{out}}},
    \label{eq:outflow_rate}
\end{equation}
where $v_{\mathrm{out}}=FWHM_{\mathrm{broad}}/2 + |v_{\mathrm{broad}}-v_{\mathrm{narrow}}|$ is the outflow velocity (with $FWHM_{\mathrm{broad}}$ the full width at half maximum of the broad component, while $v_{\mathrm{broad}}$ and $v_{\mathrm{narrow}}$ the velocity peaks of the broad and narrow components, respectively; \citealt{Rupke05a}), and $R_{\mathrm{out}}$ is the outflow radius as obtained in Sect. \ref{subsec:cube_stacking}. This model is consistent with a constant outflow rate over time. However, different outflow histories and geometries could also be adopted, leading to different results. For instance, a spherical or multi-conical geometry (i.e., a decaying outflow history) can provide outflow rates up to three times larger than found with Eq. \ref{eq:outflow_rate} (e.g., \citealt{Maiolino12,Cicone14}), although this seems to be disfavored by many observations of local galaxies (see \citealt{Lutz20} and references therein). With this caveat in mind, we obtained the outflow properties reported in Table \ref{tab:outflow_prop}, along with their corresponding mass-loading factors.

It is worth noting that the outflow velocities of the cold gas found in our work are comparable with those obtained from [CII]-based studies in normal SFGs (e.g., \citealt{Gallerani18,Ginolfi20,Herrera21}). Such velocities could be slightly larger than what found through measurements of absorption lines (e.g., NaD$\lambda\lambda~5890,5896~$\AA, OH) tracing the cold gas (e.g., \citealt{Cazzoli16,Janssen16,Roberts19}; but see also \citealt{Veilleux05,Heckman17} whose absorption line studies revealed the presence of fast ($\gtrsim500~$km/s) cool outflowing gas in different systems), and sometimes (i.e., for $v{_\mathrm{out}}\gtrsim250~\mathrm{km~s^{-1}}$) not predicted by numerical or hydrodynamic simulations (e.g., \citealt{Kim20,Andersson23}). On the other hand, \cite{Scannapieco17} used a suite of three-dimensional simulations to reproduce the evolution of initially hot material (typically quite fast and highly ionized) ejected by starburst-driven galactic outflows suggesting that, an explanation for the different velocity range of cold outflowing gas found in observations, could be that absorption lines probe the cold gas at the smallest radii of a galaxy, while emission lines trace cold material condensed from an initial hot medium at larger distances\footnote{In this regard, the [CII] line has already been proved to trace large scale cold gas emission around galaxies (e.g., \citealt{Fujimoto20,Ginolfi20}).}. \cite{Schneider18} made use of a suite of high-resolution isolated galaxy models to investigate the origin of fast-moving cool gas in outflows. They found that such gas can originate from a rapid cooling of the hot gas phase, that can generate cool gas outflows at velocities up to $\sim1000~\mathrm{km~s^{-1}}$. Interestingly, \cite{Pizzati20} used semi-analytical models to simulate [CII] emission from supernova-driven cooling outflows. Similarly to \cite{Scannapieco17}, they predicted that gas can cool very rapidly within the central kpc of the galaxies, so to guarantee the formation and survival of [CII] ions in the outflows. Particularly, they found that [CII] can be transported by the neutral outflows at velocities of $300-500~\mathrm{km~s^{-1}}$ (as found in this work and previous observations, e.g., \citealt{Ginolfi20}), likely producing the extended [CII] halos observed around high-$z$ (and possibly local; see Sect. \ref{subsec:cube_stacking} and Fig. \ref{fig:channel_maps}) galaxies (e.g., \citealt{Fujimoto20}). 
%An in-depth investigation of the origin of the cool gas entrained by outflows is beyond the scope of this work. 
Future comparisons between observations and tailored simulations will hopefully allow us to fully characterize the multi-phase nature of galactic outflows.

We show in Fig. \ref{fig:outflow_sfr} the atomic outflow rate as a function of the SFR as obtained from the spectral stacking of our galaxies and from individual detections of the broad component, and color-coded by their mass-loading factors. We also report the best-fit relations between molecular outflow rate and SFR for both local AGNs and starburst/SFGs as found by \cite{Fluetsch19}\footnote{As reported in \cite{Fluetsch19}, the ionized, neutral, and molecular phase of the ISM could contribute at the same level to the outflow rate. Therefore, we point out that the comparison between the atomic and molecular outflow rates in Fig. \ref{fig:outflow_sfr} is reasonable, as we are probing the contribution of one ISM phase in both cases.}. AGN hosts are characterized by $\eta\gg1$ in the range of SFR spanned by our sample, while SFGs have typically lower outflow efficiencies. Most of our galaxies lie along the 1:1 relation (i.e., $\eta=1$) as also found in previous observations of local SFGs (e.g., \citealt{Cicone14,Fluetsch19}), with an average mass-loading factor $\eta\sim1.3$. We note that, if we assume that all the phases of the ISM contribute equally to the outflow rate (e.g., \citealt{Fluetsch19}), we could obtain an average outflow efficiency three times larger than estimated (i.e., $\eta\gtrsim3$). From the stacking we found similar results, that is $\eta=0.97$ for the whole sample and $\eta=1.76$ for the non-detected outflows only, as obtained by assuming the corresponding median SFRs of the two sub-samples, i.e., 0.22 and 0.06 $\mathrm{M_{\odot}~yr^{-1}}$, respectively. The best fit to the individual outflow detections provided $\mathrm{log(\dot{M}_{out})=1.13~log(SFR_{[CII]})+0.07}$, with the slope in agreement with that found by \cite{Fluetsch19} for local SFGs, but closer to the 1:1 relation.     

%\begin{landscape}
%\begin{table}
\begin{sidewaystable*}
\centering
\caption{Outflow properties.}
\begin{center}
\begin{spacing}{1.25}
\begin{tabular}{l c c c c c c c c c c c c}
\hline
\hline
\multicolumn{12}{c}{Individual outflow detections}\\\\
Source & $\mathrm{v_{out}}$ & $\mathrm{v_{esc}}$ & $\mathrm{r_{halo}}$ & $\mathrm{log(M_{halo})}$ & $\mathrm{log(M_{out})}$ & $\mathrm{log(\dot{M}_{\mathrm{out}})}$ & $\eta_{\mathrm{[CII]}}$ & $\mathrm{log(\tau_{dep,out})}$ & $\mathrm{log(\dot{E}_{out})}$ & $\mathrm{log(\dot{P}_{out})}$ & $\mathrm{f_{esc}}$\\
 & [km~s$^{-1}$] & [km~s$^{-1}$] & [kpc] & $\mathrm{[M_{\odot}]}$ & $\mathrm{[M_{\odot}]}$ & $\mathrm{[M_{\odot}~yr^{-1}]}$ & & [yr] & [erg~s$^{-1}$] & [g~cm~s$^{-2}$] & [\%]\\
 (1) & (2) & (3) & (4) & (5) & (6) & (7) & (8) & (9) & (10) & (11) & (12)\\
\hline
Haro2 & $780\pm140$ & $174\pm12$ & $74\pm3$ & $10.66^{+0.05}_{-0.06}$ & $6.10^{+0.12}_{-0.16}$ & $0.01^{+0.13}_{-0.19}$ & 1.14 & $9.00^{+0.02}_{-0.03}$ & $41.29^{+0.13}_{-0.19}$ & $33.70^{+0.15}_{-0.22}$ & $62\pm3$\\
Haro3 & $153\pm34$ & $210\pm9$ & $87\pm2$ & $10.88^{+0.03}_{-0.03}$ & $6.19^{+0.18}_{-0.32}$ & $-0.62^{+0.20}_{-0.37}$ & 0.53 & $9.27^{+0.06}_{-0.08}$ & $39.25^{+0.16}_{-0.26}$ & $32.37^{+0.21}_{-0.41}$ & $18\pm4$\\
Haro11 & $314\pm32$ & $276\pm11$ & $111\pm2$ & $11.20^{+0.03}_{-0.03}$ & $7.57^{+0.12}_{-0.16}$ & $1.08^{+0.12}_{-0.17}$ & 1.75 & $9.00^{+0.04}_{-0.04}$ & $41.57^{+0.08}_{-0.10}$ & $34.38^{+0.13}_{-0.18}$ & $23\pm2$\\
He2-10 & $290\pm18$ & $238\pm10$ & $98\pm3$ & $11.03^{+0.04}_{-0.04}$ & $6.64^{+0.09}_{-0.11}$ & $0.12^{+0.09}_{-0.12}$ & 1.09 & $9.05^{+0.07}_{-0.08}$ & $40.54^{+0.05}_{-0.06}$ & $33.38^{+0.09}_{-0.12}$ & $32\pm2$\\
HS1222 & $281\pm106$ & $288\pm6$ & $114\pm1$ & $11.25^{+0.01}_{-0.01}$ & $6.17^{+0.23}_{-0.53}$ & $-0.37^{+0.25}_{-0.70}$ & 1.95 & $8.64^{+0.10}_{-0.14}$ & $40.02^{+0.24}_{-0.61}$ & $32.88^{+0.27}_{-0.93}$ & $35\pm16$\\ 
Mrk930 & $262\pm108$ & $248\pm14$ & $101\pm3$ & $11.07^{+0.05}_{-0.05}$ & $6.82^{+0.30}_{-2.20}$ & $0.25^{+0.32}_{-1.06}$ & 0.75 & $9.27^{+0.22}_{-0.48}$ & $40.59^{+0.26}_{-0.76}$ & $33.47^{+0.34}_{-0.79}$ & $24\pm5$\\
Mrk1089 & $287\pm32$ & $225\pm13$ & $92\pm3$ & $10.96^{+0.04}_{-0.04}$ & $7.37^{+0.12}_{-0.16}$ & $0.83^{+0.12}_{-0.17}$ & 1.62 & $8.99^{+0.04}_{-0.04}$ & $41.25^{+0.09}_{-0.11}$ & $34.09^{+0.13}_{-0.18}$ & $33\pm3$\\ 
UGC4483 & $439\pm130$ & $68\pm2$ & $32\pm1$ & $9.55^{+0.03}_{-0.03}$ & $2.96^{+0.25}_{-0.67}$ & $-3.39^{+0.26}_{-0.79}$ & 0.23 & $9.13^{+0.15}_{-0.23}$ & $37.39^{+0.20}_{-0.39}$ & $30.05^{+0.28}_{-0.96}$ & $95\pm2$\\ 
UM311 & $274\pm49$ & $289\pm20$ & $116\pm5$ & $11.26^{+0.06}_{-0.07}$ & $6.18^{+0.16}_{-0.26}$ & $-0.37^{+0.17}_{-0.29}$ & 0.59 & $9.27^{+0.03}_{-0.03}$ & $40.00^{+0.13}_{-0.19}$ & $32.86^{+0.18}_{-0.32}$ & $19\pm3$\\ 
UM448 & $453\pm64$ & $328\pm11$ & $129\pm3$ & $11.40^{+0.03}_{-0.03}$ & $7.46^{+0.11}_{-0.16}$ & $1.13^{+0.13}_{-0.18}$ & 1.85 & $8.98^{+0.03}_{-0.04}$ & $41.94^{+0.11}_{-0.14}$ & $34.58^{+0.13}_{-0.20}$ & $40\pm2$\\ 
UM461 & $724\pm212$ & $123\pm12$ & $54\pm4$ & $10.25^{+0.07}_{-0.08}$ & $4.68^{+0.18}_{-0.32}$ & $-1.45^{+0.20}_{-0.40}$ & 2.38 & $8.30^{+0.07}_{-0.09}$ & $39.77^{+0.20}_{-0.38}$ & $32.21^{+0.22}_{-0.48}$ & $86\pm4$\\ 
\hline
\multicolumn{12}{c}{Stacking}\\\\
Whole & $219\pm27$ & - & - & - & $5.97^{+0.13}_{-0.19}$ & $-0.68^{+0.13}_{-0.18}$ & 0.97 & - & $39.50^{+0.10}_{-0.12}$ & $32.46^{+0.14}_{-0.20}$ & -\\
Non-det. & $299\pm50$ & - & - & - & $5.52^{+0.16}_{-0.26}$ & $-0.99^{+0.16}_{-0.26}$ & 1.76 & - & $39.46^{+0.13}_{-0.18}$ & $32.28^{+0.17}_{-0.28}$ & -\\
\hline
\end{tabular}
\end{spacing}
\tablefoot{Column description: (1) source name; (2) outflow velocity; (3) escape velocity; (4) virial radius; (5) halo mass; (6) mass of the outflowing atomic gas; (7) atomic mass outflow rate; (8) mass-loading factor; (9) outflow depletion timescale; (10) outflow kinetic power; (11) outflow momentum rate; (12) outflow escape fraction. In the bottom part of the table, first column refers to the sample used for stacking. We do not report the uncertainties on the mass-loading factors as they represent lower limits (see Sect. \ref{subsec:outflow_efficiency}).}
\end{center}
\label{tab:outflow_prop}
%\end{table}
%\end{landscape}
\end{sidewaystable*}

Furthermore, we compare our results to those found at high redshift by \cite{Gallerani18} and \cite{Ginolfi20}, who took advantage of [CII] emission detected in a sample of nine SFGs at $z\sim5.5$ \citep{Capak15}, and in the sample of normal SFGs at $4<z<6$ as part of the ALMA Large Program to INvestigate [CII] at Early times (ALPINE; \citealt{Bethermin20,Faisst20,LeFevre20}), respectively. Both results are in nice agreement with our low-redshift dwarf galaxies, suggesting that similar feedback mechanisms can be in place in this kind of sources. The interpretation of this is not straightforward, being the environment and physical processes ruling the formation and evolution of high-$z$ galaxies quite different from those undergoing in the local universe. Cosmological simulations predict a roughly constant mass-loading factor at different redshifts for galaxies with $\mathrm{log(M_{*}/M_{\odot})<10}$, along with an increase with lower stellar masses (e.g., \citealt{Nelson19}). However, many observations (including this work) find no significant differences in the efficiency of outflows in primordial main-sequence galaxies and local less massive sources (e.g., \citealt{Gallerani18,Ginolfi20,Calabro22}). An explanation of this can reside in the fact that local low-metallicity dwarf galaxies could be considered as analogs of high-$z$ sources, as they can share similar properties in terms of morphology, size, metal content, or specific SFR (e.g., \citealt{Patej15,Izotov21,Motino21,Henkel22,Shivaei22}). For instance, \cite{Motino21} studied the properties of a sample of 11 potential local analogs to high-$z$ galaxies, selected to have similar SEDs to those of $z\gtrsim2$ objects. They computed the star-formation histories of their sources, finding that half of the candidates (with 3 of them also included in our sample) are characterized by a lack of star-formation activity at look-back times $\gtrsim1~$Gyr (i.e., they have no old stellar populations), thus resembling early objects. This is further supported by the recent results from \cite{Shivaei22}, who constrained the infrared SEDs of $z\sim2.3$ subsolar-metallicity galaxies and compared them with those of local dwarfs (including DGS sources) and (U)LIRGs. They found that infrared SEDs of sources in their sample have much more similar properties to those of local dwarf galaxies than to the SEDs of nearby (U)LIRGs, suggesting that local dwarfs and high-$z$ galaxies could share analogous ISM ionization properties and dust populations. In addition, their galaxies present rather high specific SFRs relative to the $z\sim2$ main-sequence, as also found for some sources in our DGS sample (see Fig. \ref{fig:ms}). Following this, local dwarfs and high-$z$ galaxies could also share comparable outflow efficiencies. On the other hand, external environment may also have an impact on $\eta$. \cite{Calabro22} characterized galactic outflows by analyzing ISM absorption lines in the spectra of 330 galaxies from the VANDELS survey \citep{Pentericci18,McLure18,Garilli21} distributed over a wide redshift range, i.e., $2<z<5$. Again, they obtained an average mass-loading factor of order of unity, with no redshift evolution. Interestingly, they found evidence for a larger contribution of inflows at earlier cosmic times that, combined with the more turbulent ISM and higher merging activity (e.g., \citealt{DeBreuck14,Jones21,Romano21}), could level out the outflow efficiency in these galaxies.

\subsection{Are outflows able to escape dark matter halos?}\label{subsec:escape_vel}
Theoretical models predict that, because of their small potential wells, outflows could quite easily bring gas outside of low-mass galaxies, clearing these sources of their metal and dust content and enriching the IGM (e.g., \citealt{Dekel86,Springel03}). In order to test such predictions, we computed the escape velocities of our dwarf galaxies ($v_{esc}$), needed by outflows to escape their gravitational potential. 

Following \cite{Fluetsch19}, we assumed a Navarro-Frenk-White (NWF) dark matter density profile \citep{Navarro96}, resulting in:
\begin{equation}
    v_{esc}(r)=\sqrt{2 |\Phi(r)|} = \sqrt{\frac{2 M_{halo} G}{r(\mathrm{ln}(1+c) - c/(1+c))} \mathrm{ln}(1+r/r_s)},
    \label{eq:escape_vel}
\end{equation}
where $G$ is the gravitational constant, $c$ is the concentration parameter\footnote{Instead of assuming a single value of $c$ for each galaxy, we used the $z=0$ relation by \cite{Duffy08} to link the concentration parameter to the halo mass as $\mathrm{log}(c)=0.76 - 0.1~\mathrm{log}(M_{halo})$.}, $M_{halo}$ is the mass of the halo, and $r_s = r_{halo}/c$ is the characteristic radius, with $r_{halo}$ as the virial radius. The halo mass was obtained from the stellar mass of the corresponding galaxy through abundance-matching techniques \citep{Behroozi10}, while the virial radius is defined as (e.g., \citealt{Lokas01,Huang17}):
\begin{equation}
    r_{halo} = \left[ \frac{3 M_{halo}}{4 \pi~200~\rho_{crit,0}} \right]^{1/3},
    \label{eq:virial_radius}
\end{equation}
with $\rho_{crit,0}$ being the present critical density. All the above-mentioned parameters are listed in Table \ref{tab:outflow_prop}.

\begin{figure}
    \begin{center}
	\includegraphics[width=\columnwidth]{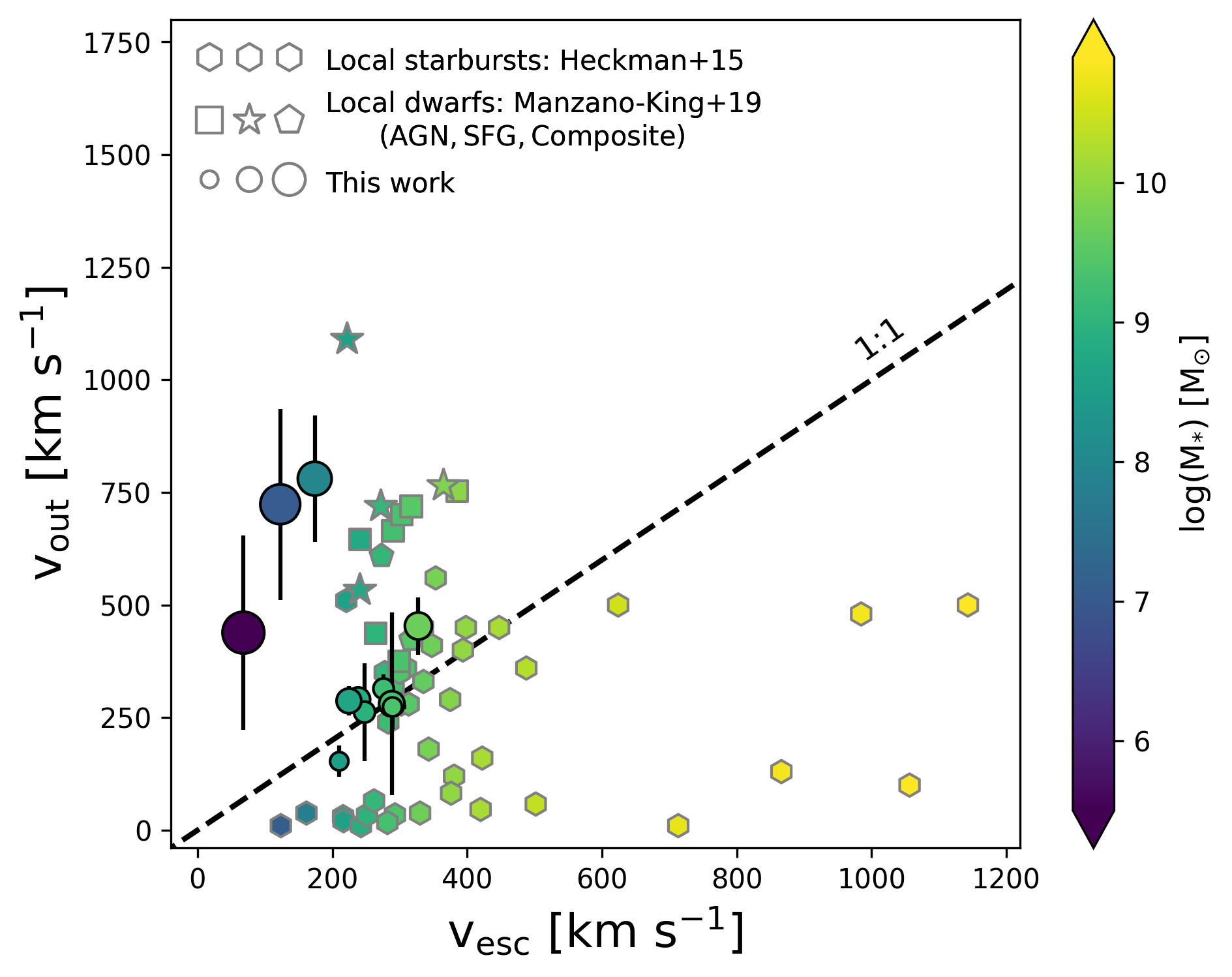}
	\end{center}
    \caption{Relation between the outflow velocity and the escape velocity. Galaxies of our sample with individual outflow detections are shown as circles with their size increasing with larger escape fractions. Local starbursts are represented as hexagons \citep{Heckman15}. Nearby dwarf galaxies, including AGNs, star forming and composite galaxies (based on their optical classification), are shown as squares, stars, and pentagons, respectively \citep{Manzano19}. The dashed line reports the 1:1 relation. All data are color-coded for their stellar mass.}
    \label{fig:escape_vel}
\end{figure}

Fig. \ref{fig:escape_vel} shows the velocity of the outflow as a function of the escape velocity for each galaxy with individual outflow detection. As a comparison, we display the results for ionized gas outflows by \cite{Heckman15} in a sample of $\sim40$ local starbursts\footnote{We computed the escape velocities for the galaxies in \cite{Heckman15} by collecting the redshift and stellar mass of each galaxy in their work, and using our Eq. \ref{eq:escape_vel}.}, and by \cite{Manzano19} from Keck spectroscopy of local dwarf galaxies (including AGNs and SFGs) drawn from the Sloan Digital Sky Survey Data Release 7 (SDSS DR7) catalog by \cite{Oh11}. The most massive galaxies (i.e., $\mathrm{log(M_{*}/M_{\odot})\gtrsim10}$) lie below the 1:1 relation at large escape velocities, implying that outflows in these sources (at least from the ionized phase) are not able to expel material outside of their dark matter halos. Conversely, all of our sources are close or above the relation, with outflow velocities higher than (or comparable to) the escape ones, in agreement with the results for ionized outflows in local dwarf galaxies by \cite{Manzano19}. This suggests that galactic winds in these objects are able to bring material at least in their CGM, having a large impact on their baryon cycle. 

To understand how much gas can be expelled out of our galaxies, we estimated their escape fractions ($f_{esc}$), defined as the fraction of the outflowing gas with velocity higher than the escape velocity. In particular, we integrated the [CII] emission of the broad component at velocities larger than $v_{esc}$, and divided by the total amount of gas carried by the outflow. We found values ranging from $\sim20$ to 90\% depending on the outflow velocity and potential well of the galaxy, with an average escape fraction of 40\%. This is more than a factor four larger than what found by \cite{Fluetsch19} in local SFGs and AGNs. However, as those authors pointed out, their sample do not include low-mass galaxies, for which a significant fraction of gas is expected to leave the galaxy and its halo enriching the IGM. Our results are instead consistent with nearby ULIRGs, as found through molecular OH-based ($\sim25\%$; \citealt{Gonzalez17}) or CO-based (15-40\%; \citealt{Pereira18,Herrera20}) analysis. In general, except for a few galaxies in our sample which are able to expel a large fraction of material directly into the IGM, the majority of the atomic gas in the outflow will remain bound to the gravitational potential of the galaxy (i.e., it will stay in the CGM), and it could be later re-accreted and used for future star formation, still contributing to the baryon cycle of its host. Interestingly, the galaxies with the larger escape fractions are also those with the lowest metallicities, suggesting that a significant fraction of metals in these sources is likely residing in (or outside) their halos, as also predicted by theoretical models (e.g., \citealt{Recchi13}).

We note here that our estimates on the escape velocity could be affected by the method used to compute the halo mass of the galaxies. For instance, \cite{Ostlin15} made use of velocity dispersion obtained from ionized emission lines to estimate the dynamical mass of Haro11, i.e., $\sim10^{11}~\mathrm{M_{\odot}}$. Their result is $\sim0.2~$dex lower than what we found from abundance matching (thus causing a lower escape velocity and allowing outflowing gas to escape more easily from the galaxy), but relies on the assumption that the observed line widths are mainly due to virial motions. However, turbulence produced by feedback driven by star formation could also have a large impact on the broadening of emission lines, significantly affecting the resulting estimates (e.g., \citealt{Green10,Moiseev12}). As the galaxies in our sample are hosting outflows, we decided to compute our halo masses based on abundance-matching methods.

Finally, we want to highlight that the total fraction of gas (atomic, molecular, and ionized) entrained outside of the galaxy by outflows could be larger than what estimated here based on [CII] emission. For instance, the warm/hot ionized outflowing gas is found to reach typically higher velocities than the cool molecular and atomic phases (e.g., \citealt{Rupke05b,Veilleux05,Heckman17}), likely producing larger escape fractions. However, different works suggest that the ionized phase contribute only minimally to the total outflow mass as compared to the other ISM phases (e.g., \citealt{Rupke13,Carniani15,Fluetsch19,Ramos19}), lowering its significance with respect to the cool gas traced in this work. Molecular gas is instead a fundamental component of the mass and energy budget of the outflow, as it could contribute up to 50\% to the total mass outflow rate (e.g., \citealt{Fluetsch19}). Therefore, multi-phase studies of the outflowing gas in these galaxies are needed to carefully describe the relative contribution of each outflow phase to the CGM enrichment and baryon cycle.

\begin{figure}
    \begin{center}
	\includegraphics[width=\columnwidth]{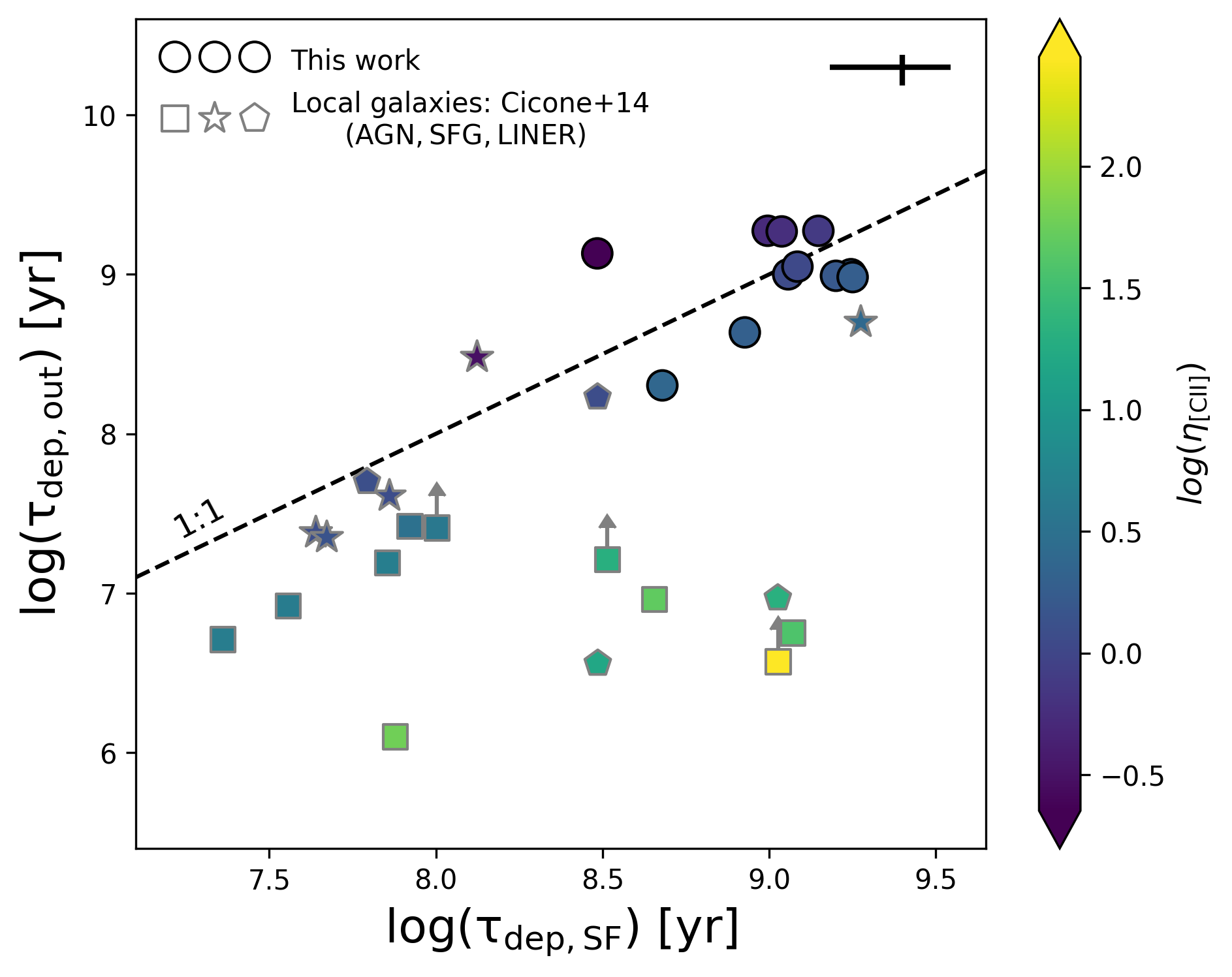}
	\end{center}
    \caption{Comparison between the depletion timescale due to outflows and the gas consumption due to star formation. DGS galaxies from this work are shown as circles. The typical uncertainty on both timescales is shown on the top-right corner. Squares, stars, and pentagons are the results from local AGNs, starburst galaxies, and LINERs, respectively \citep{Cicone14}. The dashed line reports the 1:1 relation. The area below this line is populated by galaxies whose outflows are more efficient than star formation in consuming gas (i.e., $\mathrm{\tau_{dep,out}}<\mathrm{\tau_{dep,SF}}$). All data are color-coded for their mass-loading factors.}
    \label{fig:depl_time}
\end{figure}

\subsection{Depletion timescales}\label{subsec:depletion_time}
In this section, we compare the depletion timescale due to outflows with that due to gas consumption by star formation. We define these timescales as the time needed for the molecular gas inside the galaxy to be swept out by the outflow or consumed by star formation, respectively, providing that both the outflow rate and SFR are constant over time and that no supply of fresh gas is in place e.g., via merging or cold accretion. Under these assumptions, $\tau_\mathrm{dep,out}=M_\mathrm{H2,TOT}/\dot{M}_\mathrm{out}$ and $\tau_\mathrm{dep,SF}=M_\mathrm{H2,TOT}/\mathrm{SFR}$. Here, $M_\mathrm{H2,TOT}$ is the \textit{total} mass of molecular gas derived from the luminosity of the narrow component of the [CII] line (see Table \ref{tab:technical}) by following \cite{Madden20}, i.e., $M_\mathrm{H2,TOT}=10^{2.12}\times{(L_\mathrm{[CII],narrow})}^{0.97}$. In particular, this relation derives from the application of the spectral synthesis code Cloudy \citep{Ferland17} to the DGS sample, and from the finding that a large fraction (70 to 100\%) of the H$_2$ mass is not traced by the CO(1-0) transition (usually considered to deduce the total molecular hydrogen) in dwarf galaxies (i.e., CO-dark gas mass), but it is well traced by [CII].

\begin{figure*}
    \begin{center}
	\includegraphics[width=0.8\textwidth]{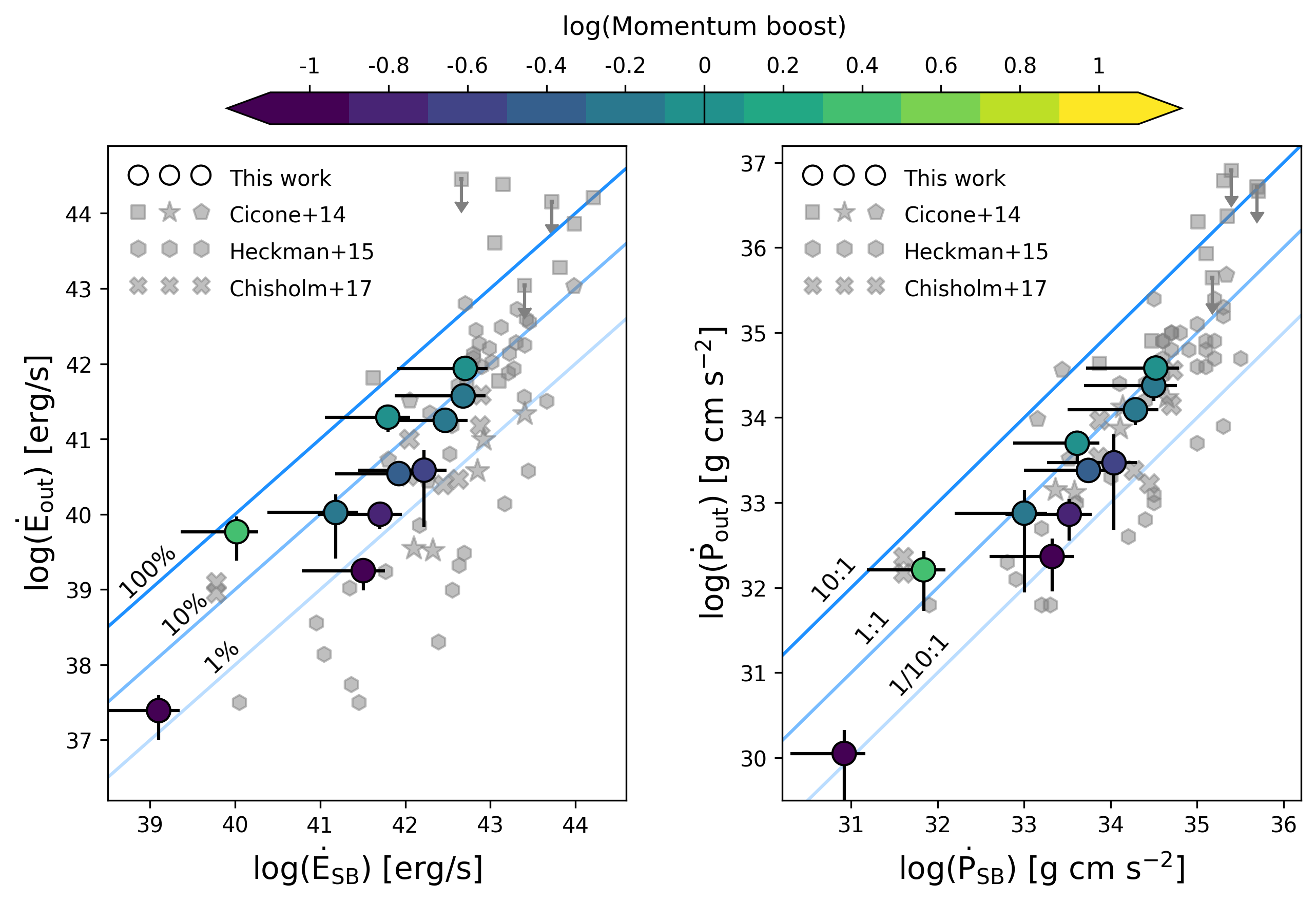}
	\end{center}
    \caption{\textit{Left:} kinetic power of the outflow as a function of the kinetic power generated by SN-driven winds. The solid lines report the relations $\dot{E}_{\mathrm{out}}=0.01, 0.1, 1~\dot{E}\mathrm{_{SB}}$, as indicated in the figure. \textit{Right:} outflow momentum rate as a function of the momentum supplied by starburst galaxies. The solid lines refer to $\dot{P}\mathrm{_{out}}=0.1, 1, 10~\dot{P}_{\mathrm{SB}}$, as reported in the figure. Both panels also show the results by \cite{Cicone14}, \cite{Heckman15}, and \cite{Chisholm17} for local starburst (stars, hexagons, and crosses) and AGN-dominated (squares and pentagons) sources. Big circles display the results from this work and are color-coded by the momentum boost, as described in the text.}
    \label{fig:energetic}
\end{figure*}

Fig. \ref{fig:depl_time} shows the relation between the two depletion timescales for our dwarf galaxies with detected outflow. Their typical error bar is displayed on the top-right corner of the figure. This was computed by propagating the errors of $M_\mathrm{H2,TOT}$ and $\dot{M}_\mathrm{out}$ (SFR) on $\tau_\mathrm{dep,out}$ ($\tau_\mathrm{dep,SF}$), and by considering that the quantities involved in the computation of both timescales are correlated to each other. The uncertainty on the molecular gas mass was instead estimated with the error propagation on the equation by \cite{Madden20} and by assuming a standard deviation of 0.14~dex on that relation, as reported in their work. In the same figure, we also compare our results with a compilation of local sources including AGNs, LINERs and starburst galaxies analyzed by \cite{Cicone14}. They found molecular outflow depletion timescales ranging from a few up to a few hundred million years for AGN-host and starburst galaxies\footnote{We caution the reader that the results found by \cite{Cicone14} are obtained under the assumption of a spherical or multi-conical geometry. Adapting their values to the outflow geometry adopted in this paper would result in outflow depletion timescales three times larger, and conversely for the mass-loading factors and for the outflow properties derived in Sect. \ref{subsec:energetic}.}, respectively, with the former populating the area with depletion timescales due to outflows much shorter than those due to star formation. In our sample, we find that the two timescales are similar, in agreement with the starburst-dominated galaxies by \cite{Cicone14}. In particular, the outflow depletion timescales of our dwarf sources range from hundred million years up to a few billion years, with $\sim60-90\%$ of our sample characterized by $\mathrm{\tau_{dep,out}}\lesssim\mathrm{\tau_{dep,SF}}$, as a consequence of the mass-loading factors larger than (or consistent with) unity. This could imply a fundamental role of galactic outflows in regulating star formation in dwarf galaxies. It is also worth noticing that, in our computation of the outflow depletion timescale, the outflow rate is only due to the atomic gas and it could be larger when accounting for the ionized and molecular ISM phases (e.g., \citealt{Fluetsch19}). For this reason, our values must be considered as upper limits, moving to lower $\mathrm{\tau_{dep,out}}$ in case of higher $\eta$, and strengthening the importance of feedback in such sources.

\subsection{Outflow energetics}\label{subsec:energetic}
Stellar winds can inject a significant amount of mechanical energy and momentum in the ISM of starburst galaxies, producing shocks which propagate outwards sweeping away the gas. Depending on the efficiency of radiative losses during this process, we can distinguish between "energy-driven" and "momentum-driven" outflows. The former are associated with adiabatic expansion of the gas powered by SN explosions, with shocks preserving their mechanical luminosity. On the other hand, radiative cooling is significant in momentum-driven outflows which are typically induced by the radiation pressure on dust grains produced by young stellar populations (e.g., \citealt{Murray05,Faucher12,Hopkins12,Cote15,Thompson15}).   

To investigate the driving mechanisms of the outflows in our galaxies, we computed their kinetic power ($\dot{E}_{\mathrm{out}}$) and momentum rate ($\dot{P}_{\mathrm{out}}$) as follows:  
\begin{equation}
    \dot{E}\mathrm{_{out}} = \frac{1}{2} \dot{M}\mathrm{_{out}} {v^2}\mathrm{_{out}},
    \label{eq:kin_power}
\end{equation}
\begin{equation}
    \dot{P}\mathrm{_{out}} = \dot{M}\mathrm{_{out}} {v}\mathrm{_{out}}.
    \label{eq:momentum}
\end{equation}
We compared these quantities with the momentum and kinetic energy supplied by starburst-driven winds, i.e., ($\dot{P}_{\mathrm{SB}}$) and ($\dot{E}_{\mathrm{SB}}$), respectively. In particular, \cite{Veilleux05} used evolutionary models of the populations of massive stars (\textit{Starburst99}; \citealt{Leitherer99}) to calculate the power injected by SN into the ISM of starburst galaxies, finding $\dot{E}\mathrm{_{SB} (erg~s^{-1})} = 7\times10^{41}~\mathrm{SFR} \mathrm{(M_{\odot}~yr^{-1})}$. Similarly, the total momentum supplied from starbursts (with the wind driven by a combination of massive star ejecta and radiation pressure) can be obtained as $\dot{P}\mathrm{_{SB} (g~cm~s^{-2})} = 4.6\times10^{33}~\mathrm{SFR} \mathrm{(M_{\odot}~yr^{-1})}$, following e.g., \citealt{Heckman15,Xu22}. As a check, we computed the total radiative momentum for our galaxies as $L_{\mathrm{bol}}/c$, where $L_{\mathrm{bol}}$ is the bolometric luminosity given by the sum of the stellar and dust luminosities from CIGALE, finding a good agreement with $\dot{P}_{\mathrm{SB}}$. The outflow kinetic power and momentum rate are reported in Table \ref{tab:outflow_prop}.

We show in Fig. \ref{fig:energetic} the relation between the kinetic energy (left panel) and momentum (right panel) carried by outflows and the corresponding quantities provided by starbursts. Points are color-coded by the "momentum boost" of each source, defined as the ratio between the outflow momentum and the radiative momentum of the galaxy (i.e., $\dot{P}_{\mathrm{out}}$/$\dot{P}_{\mathrm{SB}}$). Most of our galaxies are characterized by kinetic powers of the outflows between 1\% and 20\% of the kinetic power produced by SN, and momentum rates comparable with (or lower than) those supplied by starbursts. 

Our findings are in good agreement with previous results from the literature for ionized and molecular outflows in local galaxies. For instance, \cite{Heckman15} analyzed the properties of ionized outflows in a sample of $\sim40$ low-z starburst galaxies through UV absorption lines. We show their results in the right panel of Fig. \ref{fig:energetic} for the momentum rate, and we derived our own values of the kinetic power based on their estimates of SFR, ${v}\mathrm{_{out}}$, and $\dot{M}\mathrm{_{out}}$, as done for our sources. In their work, they found a net distinction between stronger ($\dot{P}\mathrm{_{out}} \gtrsim 10^{34}~\mathrm{g~cm~s^{-2}}$) and weaker outflows, with the former usually carrying a larger fraction of the starburst momentum (mostly lying along the 1:1 relation), and ultimately suggesting a momentum-driven outflow scenario. Similarly, \cite{Cicone14} studied the properties of molecular outflows in a sample of $\sim20$ local starburst and AGN-host galaxies. We report their results in Fig. \ref{fig:energetic}. They found large momentum boosts ($\gtrsim10$) and kinetic powers for AGN-dominated sources, requiring energy-conserving mechanisms. On the other hand, starbursts are characterized by outflow kinetic power corresponding to only few percents of that produced by SN, and momentum rate comparable to $L_{\mathrm{bol}}/c$, from which they conclude that molecular outflows in these sources are mostly momentum-driven. Finally, we also show the results by \cite{Chisholm17} who constrained the energetic of ionized outflows in a sample of eight local SFGs. They found that 1 up to 20 per cent of the energy released by SN is converted into kinetic energy of the outflow, and that outflows are more efficient in galaxies with lower stellar mass, for which additional sources of momentum apart from SN are needed.

Our galaxies populate the parameter space covered by these previous works, although with some scatter. The computed kinetic powers show that only a few galaxies would require large coupling efficiencies, of the order of 20\%, in order for their outflows to be driven by SN, while the objects with the lowest momentum boost have outflow kinetic power in agreement with a SN-driven wind with coupling efficiencies below 10\%, as predicted by models (e.g., \citealt{Efstathiou00,Veilleux05,Harrison14,Hu19}). At the same time, the momentum budget shows that most of our galaxies are comparable to $\dot{P}_{\mathrm{SB}}$, suggesting that radiation pressure on dusty clouds can contribute significantly to (or even dominate) the outflow powering mechanism, supporting the momentum-driven scenario (e.g., \citealt{Cicone14,Costa18}).

Finally, we note that other feedback mechanisms can be invoked in order to explain the outflow energetic. For instance, stellar activity can produce cavities of hot gas in the ISM of galaxies. SN can inject energy into these cavities, producing bubble-like structures that can expand outwards. Eventually, these bubbles can merge together into super-bubbles that can reach the edge of the gaseous disk of a galaxy and breakout, forming an outflow and ejecting a large fraction of metals into the CGM (e.g., \citealt{Recchi13,Cote15,Hu19}). Plenty of ionized and molecular gas observations have evidenced the presence of super-bubbles into the ISM of local galaxies, including a few DGS sources (e.g., \citealt{Cresci17,McQuinn18,McCormick18,Suzuki18,Menacho19,McQuinn19,Watkins23}). These observations are supported by as many analytical models and hydrodynamic simulations trying to describe the evolution and propagation of super-bubbles across the galaxy ISM (e.g., \citealt{Veilleux05,Faucher12,Cote15,Kim17,Fielding18,Oku22,Orr22}). Assuming typical properties of the bubbles observed in local galaxies (e.g., \citealt{Menacho19,McCormick18,Watkins23}), these models predict momentum rates in the range $10^{31}$-$10^{35}$~g~cm~s$^{-2}$ (e.g., \citealt{Veilleux05,Orr22}), comparable with what found for our outflows and in agreement with the typical mechanical momentum released by star formation and SN in starburst-driven winds (see Fig. \ref{fig:energetic}, right panel). This suggests that SN super-bubbles could be an important channel for the origin of galactic outflows in dwarf galaxies and for the CGM/IGM enrichment. However, such estimates are based on many assumptions on the e.g., size and velocity of the expanding bubbles, SN rate, or ambient number density (e.g., \citealt{Veilleux05,Kim20}), and can be affected by large uncertainties, preventing us from firm conclusions.

All of the feedback processes introduced in this section could jointly concur to the winds powering mechanism. Therefore, additional data and increased statistics are needed to properly constrain their individual contribution to the outflow energetic.

%Darko's paper:\\
%Check dust to metal ratio as a proxy of outflow (fig.8)\\
%Check quenching time with respect to depletion time (fig.6)\\

%Energetic considerations and comparison:\\
%\cite{Calabro22}, \cite{Molina22}, \cite{Harrison18}, \cite{Chisholm17}, %\cite{Garcia15} (also for radius estimation)\\

%\cite{Marasco22} for comparison with dwarfs.

%%%%%%%%%%%%%%%%%%%%%%%%%%%%%%%%%%%%%%%%%%%%%%%%%%%
\section{Summary and conclusions}\label{sec:summary}
Here, we investigate the impact of galactic outflows in a sample of 29 local low-metallicity dwarf galaxies drawn from the DGS \citep{Madden13,Madden14}. These sources benefit from \textit{Herschel}/PACS spectroscopic coverage of the rest-frame FIR, as well as from a collection of multi-wavelength ancillary data, which make them an ideal sample to study how feedback affects their evolution. We take advantage of [CII] observations from PACS to carry out a systematic analysis of atomic outflows in these galaxies by looking at the broad wings in their spectra. Our main results are summarized as follows.
\begin{itemize}
    \item We fitted the [CII] line profiles of all the galaxies in our sample with a single-Gaussian model, and inspect the corresponding residuals searching for an excess of emission at the high-velocity tails of the spectra (suggestive of the presence of outflowing gas, see Sect. \ref{subsec:individual_out}). We found such an excess in 11 out of 29 galaxies, with three of them better described with a three-Gaussian model likely because of past or undergoing merging activity.
    \item Having excluded the null hypothesis that the average [CII] emission of our galaxies could be modeled with a single Gaussian component only (see Sect. \ref{subsec:spec_stacking}, Fig. \ref{fig:stacked_res}), we proceeded with a variance-weighted stacking of the [CII] spectra of both the whole sample of 29 dwarf galaxies, and the sub-sample of 18 sources with no individual wings detection, in order to include the remaining galaxies with no or weak outflow emission (Fig. \ref{fig:spec_stacking}). In the two cases, we observed significant residuals at velocities $\pm\sim400~\mathrm{km~s^{-1}}$, which are reduced to the noise level by fitting the spectra with a double-Gaussian model. From the stacking of both samples, we measured average FWHM of $\sim240$ and $\sim460$~km~s$^{-1}$ for the narrow and broad component, respectively.
    \item We implemented a spatial stacking of the [CII] data cubes to measure the average size of the outflow (Sect. \ref{subsec:cube_stacking}, Fig. \ref{fig:channel_maps}). By fitting the intensity map of the stacked wings, we found a typical outflow radius of $R\mathrm{_{out}}\sim1~$kpc, in agreement with previous estimates for local galaxies (e.g., \citealt{Arribas14,Fluetsch19}). Furthermore, we found that the core of the stacked [CII] emission is significantly more extended than the typical UV size of these sources (tracing their stellar contribution), likely highlighting the presence of [CII] halos around them, as also found at higher redshift (e.g., \citealt{Fujimoto19,Ginolfi20}).    
    \item We constrained the outflow efficiency in our galaxies by providing an estimate of their mass-loading factor (i.e., by dividing their outflow mass rate for the SFR). We obtained $\eta\mathrm{_{[CII]}}\sim1$ for most of the sources, in agreement with both local and high-$z$ SFGs (e.g., \citealt{Gallerani18,Fluetsch19,Ginolfi20}), thus suggesting that the evolution of these galaxies could be ruled by similar feedback mechanisms (see Sect. \ref{subsec:outflow_efficiency}, Fig. \ref{fig:outflow_sfr}).
    \item We compared the outflow velocities with the escape velocities needed by galactic winds to bring gas outside of their host halos (Sect. \ref{subsec:escape_vel}, Fig. \ref{fig:escape_vel}). We found that, despite the low mass-loading factors, galactic outflows are able to sweep ISM material out of dwarf galaxies (i.e., directly into the IGM), with an average fraction of expelled atomic gas of 40\%, that increases up to $\sim90\%$ for very low-mass and low-metallicity sources.
    \item Assuming constant outflow rate and SFR, we derived the depletion timescales due to outflows and gas consumption by star formation, respectively (Sect. \ref{subsec:depletion_time}, Fig. \ref{fig:depl_time}). Our results cover a similar interval in both timescales, ranging from hundred million to a few billion years. At least 60\% of our sample is characterized by $\mathrm{\tau_{dep,out}}\lesssim\mathrm{\tau_{dep,SF}}$, with the fraction rising to $\sim90\%$ if accounting for the other phases of the ISM (i.e., ionized and molecular). This suggests a possible main role of galactic outflows in regulating the amount of gas in dwarf galaxies and their star formation histories.
    \item We investigated the energetic of the outflows by computing their kinetic power and momentum rate (see Sect. \ref{subsec:energetic}). By comparing these quantities with the energetic expected by starburst-driven winds (e.g., \citealt{Veilleux05,Heckman15}), we found that our outflows are likely powered by the radiation pressure of young stars on dusty clouds (i.e., momentum-driven scenario), as also observed in local starbursts (e.g., \citealt{Cicone14}). However, we can not exclude that SN contribute in accelerating the outflows (i.e., energy-driven scenario), as the majority of our sources is still comparable with theoretical models assuming a coupling efficiency $\lesssim10\%$ (see Fig. \ref{fig:energetic}). Moreover, both observations and simulations suggest that SN super-bubbles can also produce galactic winds in local dwarf galaxies and (under specific assumptions on the properties and composition of the bubbles) they could account for the momentum rates observed in the outflows.
\end{itemize}
Our results highlight the importance of galactic feedback in the evolution of low-mass sources and their environment. Although they are not as much efficient as AGN-driven winds, outflows powered by star formation are still able to mold the history of (especially) dwarf galaxies, for which they can easily bring dust and gas into their CGM (with a significant fraction of them even further, i.e., into the IGM). In a forthcoming paper, we will show how our constraints on the mass-loading factor impact the description of the physical processes responsible for the formation and destruction of dust in these galaxies (Nanni et al. in prep.). Further investigation is needed in order to provide a more complete picture of the effect of feedback on local galaxies. More observations, both from archival research or via follow-ups with e.g., NOEMA, ALMA, JWST or Keck telescopes, will allow us to extend our present sample, and to collect ionized and molecular data to complement all the phases of the outflows, making an in-depth study of their interplay.

\begin{acknowledgements}
We warmly thank the referee for her/his useful comments and suggestions that nicely improved the quality of our paper. M.R. would like to thank Denis Burgarella and Katarzyna Ma\l{}ek for helpful discussions. HIPE is a joint development by the \textit{Herschel} Science Ground Segment Consortium, consisting of ESA, the NASA \textit{Herschel} Science Center, and the HIFI, PACS and SPIRE consortia. This research has made use of the NASA/IPAC Extragalactic Database (NED),
which is operated by the Jet Propulsion Laboratory, California Institute of Technology,
under contract with the National Aeronautics and Space Administration. M.R., A.N., and P.S. acknowledge support from the Narodowe Centrum Nauki (UMO-2020/38/E/ST9/00077). G.C.J. acknowledges funding from ERC Advanced Grant 789056 ``FirstGalaxies’’ under the European Union’s Horizon 2020 research and innovation programme. J. is grateful for the support from the Polish National Science Centre via grant UMO-2018/30/E/ST9/00082. D.D. acknowledges support from the National Science Center (NCN) grant SONATA (UMO-2020/39/D/ST9/00720).
\end{acknowledgements}


\bibliographystyle{aa} % style aa.bst
\bibliography{aanda.bib} % your references Yourfile.bib

\begin{appendix}

\section{Spectra extraction}\label{app:bad_spectra}
\begin{figure*}[t]
\centering
\begin{subfigure}[b]{0.9\textwidth}
   \includegraphics[width=1\linewidth]{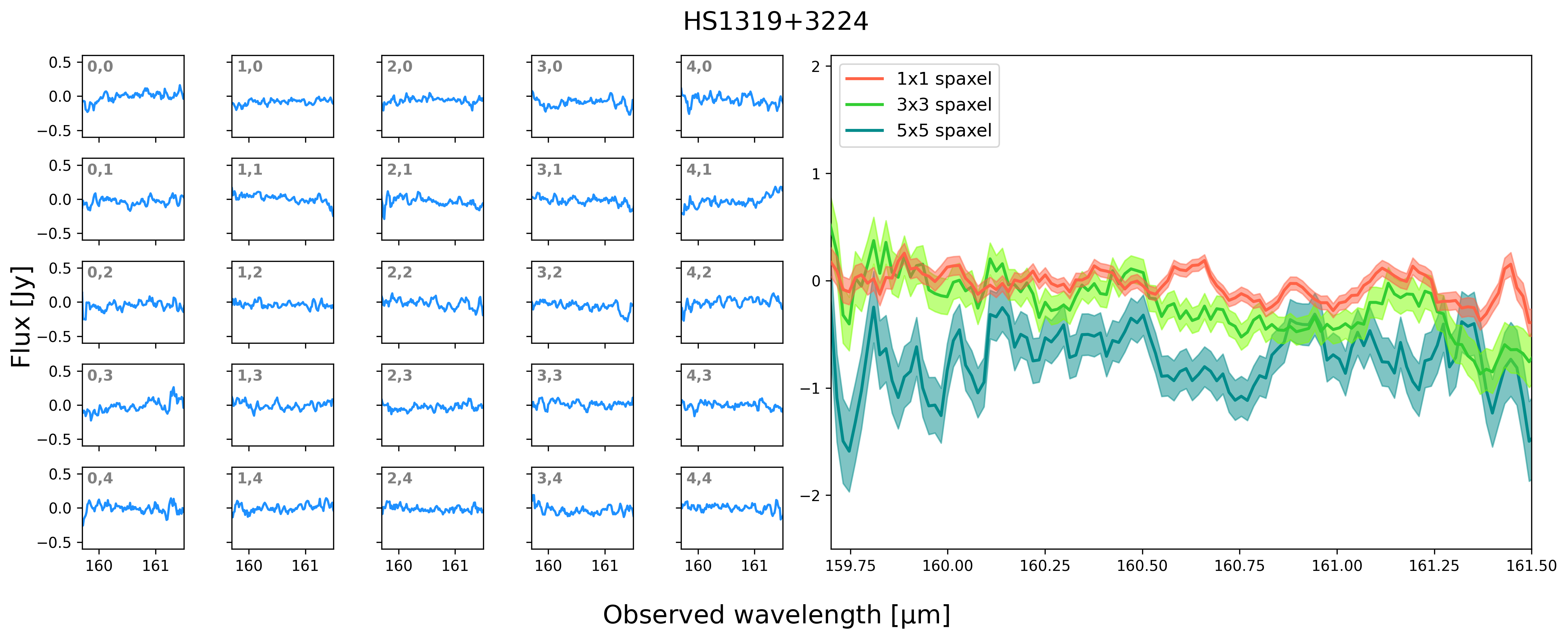}
\end{subfigure}

\vspace{0.5cm}

\begin{subfigure}[b]{0.9\textwidth}
   \includegraphics[width=1\linewidth]{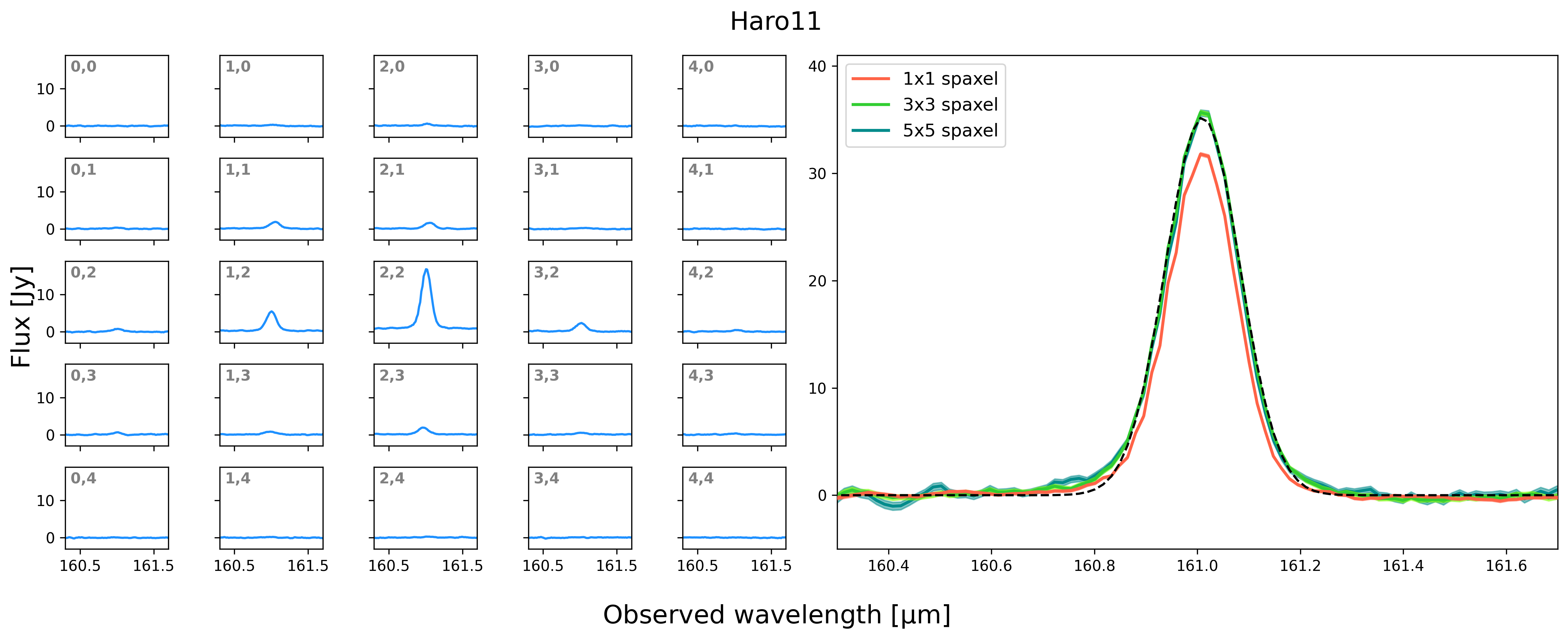}
\end{subfigure}
\caption{Examples of spectra extraction for two galaxies with bad (top) and good (bottom) quality of the [CII] data. For both galaxies: left panel shows the [CII] spectra obtained for each spaxel of the PACS data cube (with the pair numbers on the top-left of each plot associated to the corresponding position on the PACS footprint); on the right panel, the solid lines represent the spectra from the central spaxel (orange), and from the inner $3\times3$ (green) and $5\times5$ (teal) spaxel grids, while the shaded areas show their uncertainties. For Haro11, the right panel reports the continuum-subtracted spectra, while the dashed black line displays the single component Gaussian fit on the $3\times3$ inner spectrum.}\label{fig:bad_spec}
\end{figure*}

We show here two examples of spectra extraction from the PACS data cubes, as described in Sect. \ref{sec:data}. The top panels in Fig. \ref{fig:bad_spec} report the case of the galaxy HS1319+3224. On the left, the spectra retrieved from each spaxel at the position of the corresponding PACS footprint of the galaxy are shown. The spectrum from the central spaxel, and those obtained from the sum of the $3\times3$ and $5\times5$ grids are also displayed on the right. This galaxy is not present in our final sample, as it is one of the 6 sources we discarded because of their noisy [CII] spectra (see Sect. \ref{sec:data}). Indeed, by looking at the figure, none of the spectra show a clear emission feature to be used for our analysis. For comparison, we also show at the bottom of Fig. \ref{fig:bad_spec}, the [CII] spectra of Haro11, one of the 29 sources of our sample. Contrarily to HS1319+3224, this represents an example of the good quality of our data, that was not possible to reach for the above-mentioned excluded DGS galaxies. Haro11 is placed at the center of the PACS footprint, where the signal is larger than in the outskirts. The $3\times3$ inner spectrum resulted to be the one with the largest flux and S/N (see Table \ref{tab:technical}) and we used it to search for broad wings in its high-velocity tails, clearly visible in the figure as a flux excess with respect to the single Gaussian component fitted to the emission profile.   

%%%%%%%%%%%%%%%%%%%%%%%%%%%%%%%%%%%%%
\section{Individual spectral fitting}\label{app:individual_fit}
\begin{figure*}
    \includegraphics[width=.335\textwidth]{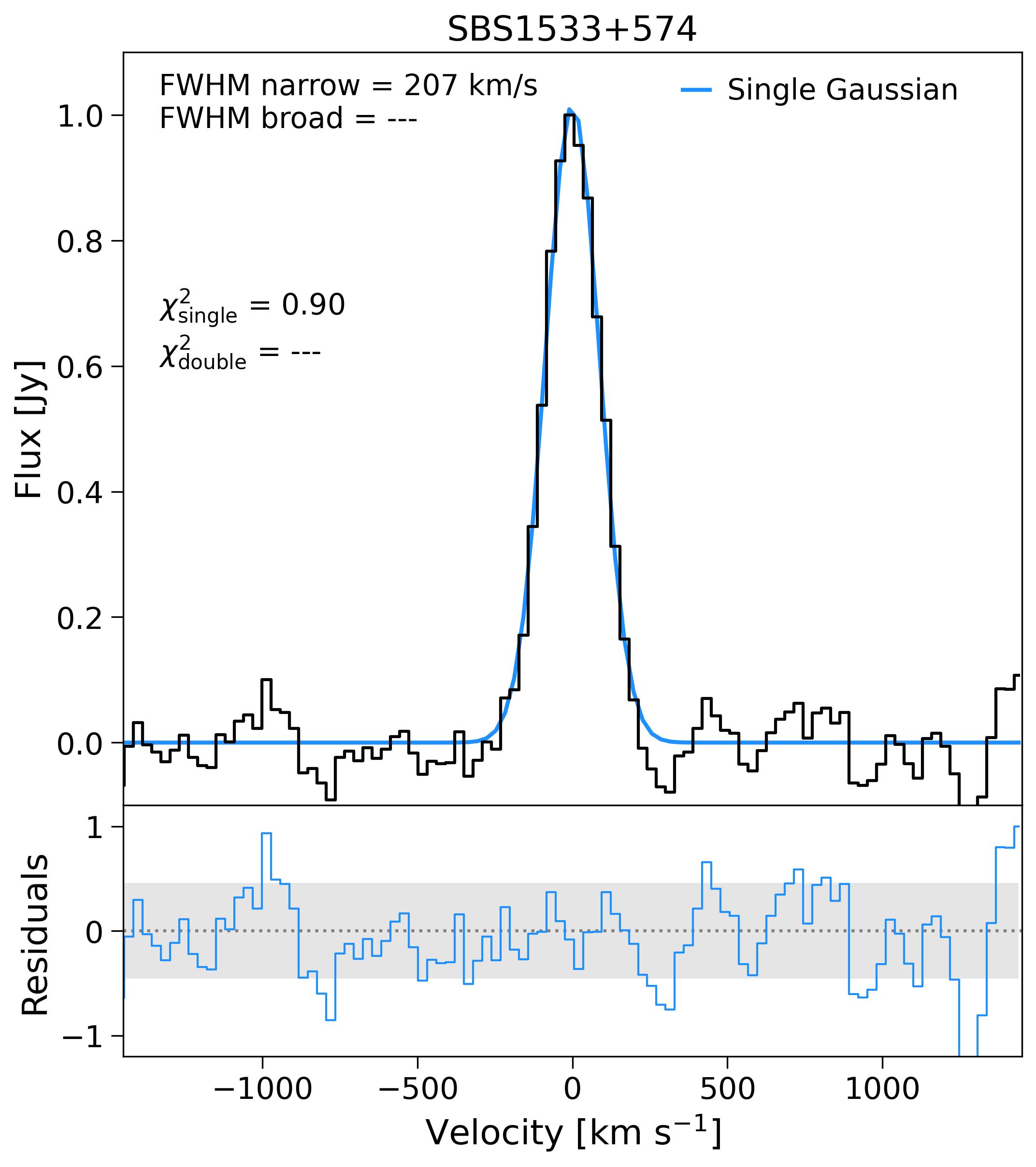}\hskip -0.5ex
    \includegraphics[width=.335\textwidth]{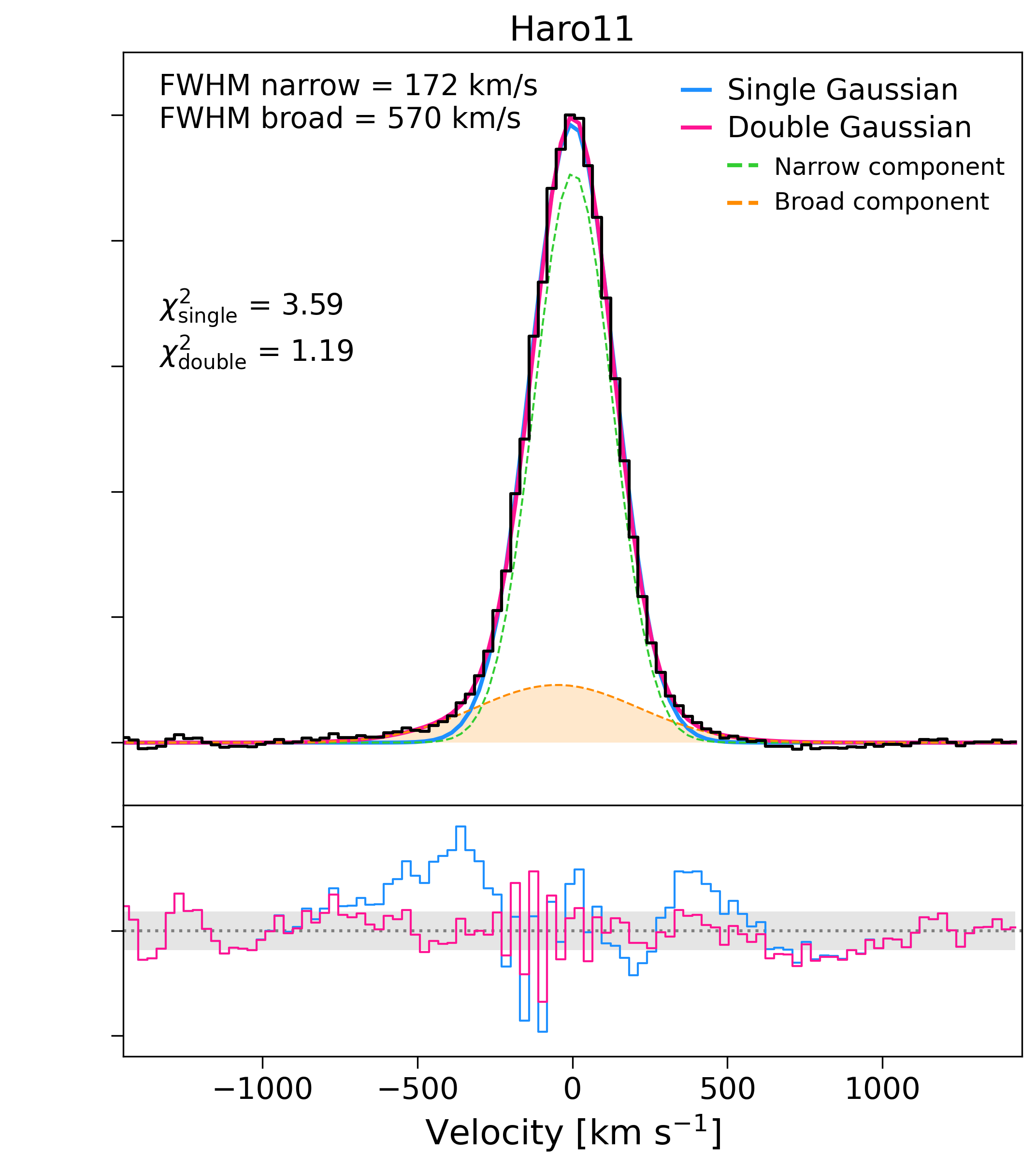}\hskip -0.5ex
    \includegraphics[width=.335\textwidth]{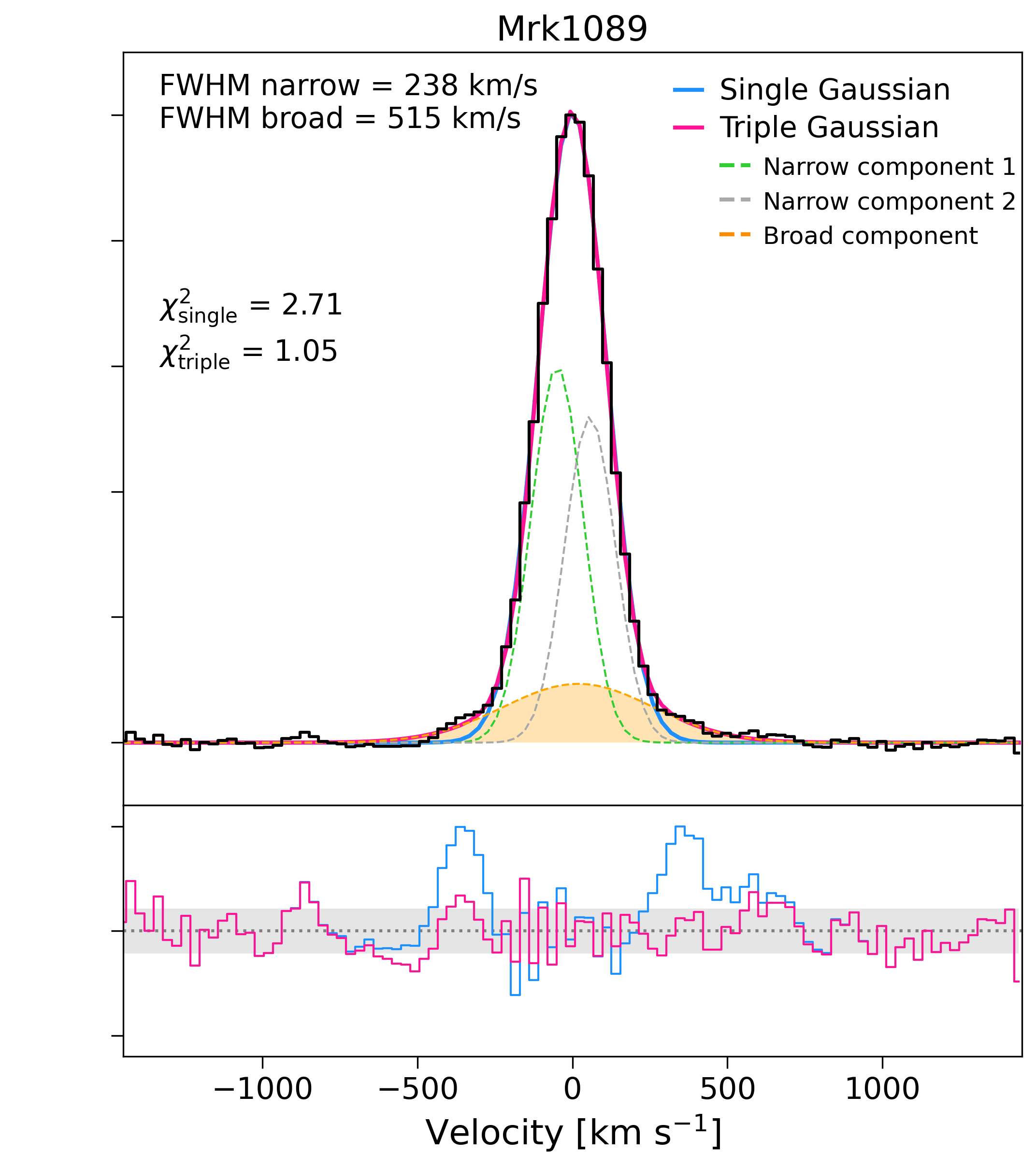}
\caption{Continuum-subtracted [CII] spectra (black histograms) as a function of the velocity offset computed with respect to the line peak. The figure shows three different galaxies with no detection of a broad component (left panel), and with individual outflow detection in the case of a single source (middle panel) and a possible merger (right panel). For each galaxy (whose name is reported at the top of the figure), the line profile is fitted with a single Gaussian function (in blue). A double (triple) Gaussian profile (in pink), that is the sum of one (green line) or two (green and grey lines) narrow and a broad (orange line) components, is also shown in the middle and right panels. The FWHM of both components and the corresponding reduced $\chi^2$ are reported in each figure. For Mrk1089, $\mathrm{FWHM_{narrow}}$ is obtained from the Gaussian profile resulting from the sum of the two narrow components. The bottom panels displays the residuals from the single (blue) and double/triple (pink) Gaussian functions. The dotted horizontal line marks the zero level, while the shaded area represents the rms of each spectrum at $\pm1\sigma$. Flux densities and residuals are both normalized to the corresponding maximum values.}\label{fig:fitting_example}
\end{figure*}  

In this appendix, we report three examples of spectral fitting to individual galaxies, as discussed in Sect. \ref{subsec:individual_out}. In the left panel of Fig. \ref{fig:fitting_example}, we show the case of the galaxy SBS1533+574, that is one of the 18 sources of our sample with no evidence for a broad component in their [CII] spectra. On the other hand, the middle panel reports the [CII] spectrum of Haro11 (the same spectrum is also shown in the bottom panel of Fig. \ref{fig:bad_spec} as a green line), in which the presence of the broad component is clearly evidenced by the large residuals at $\sim\pm400~\mathrm{km~s^{-1}}$, obtained with a single-component Gaussian fit of the line. Finally, the right panel shows the case of one of the three galaxies of our sample (Mrk1089) for which we found that a three-component modeling was a better description of their [CII] line profiles. As stated in Sect. \ref{subsec:individual_out}, previous studies found evidence for possible merging activity in these objects. Following them, we fitted their spectra with two narrow and one broad components, resulting in better residuals than obtained with a double Gaussian fit only.

\end{appendix}

\end{document}